\documentclass[12pt,a4paper]{article}
\pdfoutput=1 % if submitting a pdflatex (i.e. if you have images in pdf, png or jpg format)

\usepackage{jheppub} % for details on the use of the package, please see the JHEP-author-manual

\usepackage[T1]{fontenc} % if needed
\usepackage[mathscr]{euscript}   
\usepackage{bm,hyperref,slashed} 

\title{\boldmath$CP$ violation in $\Sigma^+\to p \ell^+\ell^-$ within the standard model and beyond}

\author[a,b]{Xiao-Gang He,}
\author[a,b]{Jusak Tandean,}
\author[c]{and German Valencia}

\affiliation[a]{State Key Laboratory of Dark Matter Physics,\\Tsung-Dao Lee Institute \& School of Physics and Astronomy,\\Shanghai Jiao Tong University, Shanghai 201210, China}
\affiliation[b]{Key Laboratory for Particle Astrophysics and Cosmology (MOE)\\\& Shanghai Key Laboratory for Particle Physics and Cosmology,\\Tsung-Dao Lee Institute \& School of Physics and Astronomy,\\Shanghai Jiao Tong University, Shanghai 201210, China}
\affiliation[c]{School of Physics and Astronomy, Monash University,\\Wellington Road, Clayton, Victoria 3800, Australia}

\emailAdd{hexg@sjtu.edu.cn}
\emailAdd{jtandean@yahoo.com}
\emailAdd{german.valencia@monash.edu}

\abstract{The LHCb collaboration has recently observed the rare hyperon decay\linebreak$\Sigma^+\to p\mu^+\mu^-$. It can also measure the corresponding antihyperon channel with comparable precision and is thus in a position to extract information on $CP$ violation in this mode. Interestingly, the long-distance contributions that dominate it within the standard model provide large absorptive phases that could drive substantial $CP$ violation through interference with potential new-physics contributions.
Here we explore this possibility, finding that the decay rate asymmetry is currently allowed to be as high as tens of percent, which can be probed by LHCb in the near future.
We additionally consider the same with regard to the dielectron mode $\Sigma^+\to pe^+e^-$ as well as the related radiative one $\Sigma^+\to p\gamma$.}

\notoc

\begin{document} \parskip=3pt
\maketitle
%\flushbottom

\newpage

{\bf\hypersetup{linkcolor=blue}\tableofcontents} 
\noindent\hrulefill

\section{Introduction\label{intro}}

The LHCb collaboration in 2025 reported the first observation of the rare hyperon decay \,$\Sigma^+ {\,\to\,} p\mu^+\mu^-$\, after discovering about 237 events in their Run-2 data \cite{LHCb:2025evf}.
With the Run-1 result \cite{LHCb:2017rdd} included, the branching fraction is measured to be \cite{LHCb:2025evf} 
\begin{equation} \label{Bexp}
{\cal B}(\Sigma^+ {\,\to\,} p\mu^+\mu^-)_{\tt exp}^{} \,=\, (1.09\pm 0.17)\times 10^{-8} \,. ~~~ ~
\end{equation}
This is unchanged if averaged with the far-less-precise finding of the HyperCP experiment~\cite{HyperCP:2005mvo}.

The value in eq.\,(\ref{Bexp}) agrees well with the prediction of the standard model (SM), in particular favoring one of the eight possibilities arising from the ambiguity in the calculation of the long-distance contribution to this mode starting from the form factors of the radiative channel \,$\Sigma^+ {\,\to\,} p\gamma$\, determined from its measurements \cite{He:2005yn,Roy:2024hqg,He:2018yzu}.
The favored prediction, made in the case dubbed ``relativistic solution 4'' in ref.\,\cite{Roy:2024hqg}, is
\begin{equation}
{\cal B}(\Sigma^+ {\,\to\,} p\mu^+\mu^-)_{\tt SM}^{} \,=\, (1.2\pm 0.1)\times 10^{-8} \,. ~~~
\end{equation}
The corresponding distribution of the dimuon invariant mass is also compatible with what LHCb observed \cite{LHCb:2025evf}.

In the years ahead, empirical studies concerning other quantities of \,$\Sigma^+ {\,\to\,} p\mu^+\mu^-$\, and its antiparticle counterpart, \,$\overline\Sigma{}^- {\,\to\,} 
\overline p\mu^+\mu^-$,\, are anticipated \cite{LHCb:2025evf,He:2018yzu,AlvesJunior:2018ldo,Goudzovski:2022vbt}.
One of the most feasible observables is the difference between the branching fractions of these two modes~\cite{LHCb:2025evf,Goudzovski:2022vbt}, which would reveal $CP$ violation if established to be nonvanishing.
Here we investigate this within and beyond the SM and do likewise with the dielectron process, $\Sigma^+ {\,\to\,} pe^+e^-$.
This latter channel is thought likely to be within reach of upcoming LHCb measurements~\cite{AlvesJunior:2018ldo,Goudzovski:2022vbt} and the ongoing BESIII experiment \cite{Li:2016tlt}, as well as future efforts like PANDA \cite{PANDA:2021ozp} and the proposed Super Tau Charm Facility \cite{Achasov:2023gey,Zhang:2026pyv}.
It is worth mentioning in passing that lattice-QCD analyses of \,$\Sigma^+ {\,\to\,} p\ell^+\ell^-$\, have lately been initiated \cite{Erben:2022tdu,Erben:2025jkq}, which could furnish complementary information once they achieve competitive precision.

The aforesaid associated transition \,$\Sigma^+ {\,\to\,} p\gamma$\, may offer another access to $CP$ violation.
A search for it in this mode has recently been carried out for the first time by~BESIII, but the results were still consistent with zero \cite{BESIII:2023fhs}.
We will also look at this theoretically.

The structure of the remainder of this paper is as follows.
In section \ref{cpvobs} we begin by outlining the contributions to \,$\Sigma^+ {\,\to\,} p\ell^+\ell^-$,\, $\ell=e,\mu$,\, from the SM and from various quark operators that can appear beyond it.
Subsequently we discuss the observables on which we concentrate that are potentially sensitive to signals of $CP$ violation in \,$\Sigma^+ {\,\to\,} p\ell^+\ell^-$\, and \,$\Sigma^+ {\,\to\,} p\gamma$.\,
In section \ref{np} we examine how much these observables could be enhanced over their SM expectations, first in a model-independent approach and then in a couple of well-known scenarios of new physics.
In section \ref{concl} we give our conclusions.
An appendix contains updates on the SM predictions.

\section{\textit{CP} asymmetries and SM expectations\label{cpvobs}}

For \,$\Sigma^+ {\,\to\,} p\ell^+\ell^-$\, we work with an amplitude of the form \cite{He:2018yzu}
\begin{align} \label{MS2pll}
{\cal M}\, & =\, \bar u_p^{} \big[ \big( \tilde{\textsc a} + \gamma_5^{} \tilde{\textsc b} \big) i \sigma^{\eta\kappa}  q_\kappa^{} - \gamma^\eta \big(\tilde{\textsc c} + \gamma_5^{}\tilde{\textsc d} \big) \big] u_\Sigma^{}\, \bar u_\ell^{} \gamma_\eta^{} v_{\bar\ell} + \bar u_p^{}\gamma^\eta \big(\tilde{\textsc e} + \gamma_5^{}\tilde{\textsc f} \big) u_\Sigma^{}\, \bar u_\ell^{} \gamma_\eta^{} \gamma_5^{} v_{\bar\ell} ~~~
\nonumber \\ & ~~~ +\, \bar u_p^{} \big( \tilde{\textsc g}+\gamma_5^{}\tilde{\textsc h} \big) u_\Sigma^{}\, \bar u_\ell^{} v_{\bar\ell} + \bar u_p^{} \big( \tilde{\textsc j} + \gamma_5^{} \tilde{\textsc k} \big) u_\Sigma^{}\,
\bar u_\ell^{} \gamma_5^{} v_{\bar\ell} \,,
\end{align}
where \,$q=P_\Sigma-P_p=P_++P_-$,\, with $P_\Sigma$, $P_p$, and $P_\pm$ being the four-momenta of the~$\Sigma^+$, the proton, and $\ell^\pm$, respectively, \,$\tilde{\textsc a},\tilde{\textsc b},.$..$,\tilde{\textsc k}$\, denote quantities that can be complex and originate from the SM and possible new physics, $u_\Sigma^{}$ and $u_p$ $(u_\ell^{}$ and $v_{\bar\ell})$ are the Dirac spinors of the baryons (leptons), and \,$\sigma^{\eta\kappa} = i[\gamma^\eta,\gamma^\kappa]/2$.\,
The formula for the decay rate derived from $\cal M$ above is available from ref.\,\cite{He:2018yzu} and will not be reproduced here.
More generally, one could add tensor-tensor terms such as \,$\bar u_p\sigma^{\eta\kappa} u_\Sigma^{}\, \bar u_\ell^{} \sigma_{\eta\kappa}^{} v_{\bar\ell}^{}$\, to~eq.\,(\ref{MS2pll}), but we ignore this possibility for simplicity, as $\cal M$ can already accommodate many new-physics models.

Within the SM the amplitude receives  contributions induced by interactions at both short and long distances.
The former proceeds at the quark level from photon- and $Z$-penguin and box diagrams \cite{Inami:1980fz,Shifman:1976de,Buchalla:1995vs} with up-type quarks and the $W$-boson running around the loops.
It is described by the effective Hamiltonian \cite{Shifman:1976de,Buchalla:1995vs,Gilman:1978bg} 
\begin{equation} \label{smH}
{\cal H}_{\rm eff}^{\tt SM} \,=\, \frac{G_{\rm F}}{\sqrt2} 
\Big[ \big(V_{\!ud}^*V_{\!us}^{} z_{7V}^{} - V_{td}^*V_{ts}^{} y_{7V}^{}\big) O_{7V}^{} - V_{td}^*V_{ts}^{} y_{7A}^{} O_{7A}^{} + V_{\!ud}^*V_{\!us}^{} C_{7\gamma}^{} O_{7\gamma}^{} \Big] \,, 
\end{equation}
where $G_{\rm F}$ symbolizes the Fermi coupling constant, $V_{kl}$ stand for the elements of the Cabibbo-Kobayashi-Maskawa (CKM) matrix, $z_{7V}^{}$, $y_{7V}^{}$, $y_{7A}^{}$, and $C_{7\gamma}$ are the Wilson coefficients at QCD renormalization scales of order 1 GeV, 
\begin{align} \label{O7}
O_{7V}^{} & \,=\, 2\,\overline d\gamma^\kappa P\!_L^{} s\, \overline\ell\gamma_\kappa^{}\ell \,, \hspace{4em} O_{7A}^{} \,=\, 2\,\overline d\gamma^\kappa  P\!_L^{} s\, \overline\ell\gamma_\kappa^{}\gamma_5^{}\ell \,, &
\nonumber \\
O_{7\gamma}^{} & \,=\, \frac{\tt e}{8\pi^2}\, \overline d \sigma_{\eta\kappa}^{} \big( m_s^{} P\!_R^{} + m_d^{}  P\!_L^{} \big) s\, F^{\eta\kappa} \,,
\end{align}
with {\tt e} being the proton's electric charge, $m_{d(s)}$ the $d(s)$-quark mass, $F^{\eta\kappa}$ the photon field-strength tensor, and \,$P_{L,R}^{} = \big(1\mp\gamma_5^{}\big)/2$\, the chiral projection operators.

The long-distance piece in the amplitude is brought about by photon exchange and can be written as \cite{He:2005yn,Lyagin:1962,Bergstrom:1987wr} 
\begin{align} \label{ffabcd}
{\cal M}_{\tt SM}^{\textsc{ld}} & \,=\, 4\alpha_0^{}\pi\, G_{\rm F}^{}\, \bar u_p^{}  \bigg[ \frac{i}{q^2} \big(a+\gamma_5^{}b\big) \sigma^{\eta\kappa} q_\kappa^{} - \gamma^\eta \big(c+\gamma_5^{}d\big) \bigg] u_\Sigma^{}\, \bar u_\ell^{} \gamma_\eta^{} v_{\bar\ell}^{} \,, ~~~~~
\end{align}
where \,$\alpha_0^{} \equiv {\tt e}^2(0)/(4\pi)=1/137.036$\, is the fine-structure constant \cite{ParticleDataGroup:2024cfk} and \,$a$, $b$, $c$, and $d$\, are form factors that are functions of $q^2$.
Being complex, they supply strong-interaction phase-shifts that are significant and enter the $CP$-violating observables of interest.

The impact of new physics (NP) beyond the SM is parameterized according to the Hamiltonian 
\begin{equation} 
{\cal H}_{\rm eff}^{\tt NP} \,=\, -\raisebox{3pt}{\footnotesize$\displaystyle\sum_{\mbox{\it\scriptsize r}}$}\, C_r^{}{\cal O}_r^{} \,+\,\rm H.c. \,, ~~~ ~~
\end{equation}
where $C_r$ designate the Wilson coefficients, which are in general complex as well, and ${\cal O}_r$ represent quark-level operators that are given in the low-energy effective field theory (LEFT) by \cite{Mertens:2011ts,Bobeth:2017ecx,Geng:2021fog,Neshatpour:2022fak,Roy:2024hqg} 
\begin{align} \label{Oj}
{\cal O}_{7(7')}^{} & = \frac{\lambda_t^{}{\tt e}G_{\rm F}^{}m_s^{}}{4\sqrt2\,\pi^2}\, F_{\eta\kappa}^{}\, \overline s \sigma^{\eta\kappa} P_{L(R)}^{} d \,, &
\lambda_{\tt q}^{} & = V_{{\tt q}d}^{} V_{{\tt q}s}^* \,,
\nonumber \\
{\cal O}_{9(9')}^\ell & = \frac{\lambda_t^{}{\tt e}^2G_{\rm F}^{}}{4\sqrt2\,\pi^2}\, \overline s\gamma^\eta P_{L(R)}^{}d\, \overline\ell\gamma_\eta^{}\ell \,, &
{\cal O}_{S(S')}^\ell & = \frac{\lambda_t^{}{\tt e}^2G_{\rm F}^{}m_s^{}}{4\sqrt2\,\pi^2}\, \overline sP_{R(L)}^{}d\, \overline\ell\ell \,,
\nonumber \\
{\cal O}_{10(10')}^\ell & = \frac{\lambda_t^{}{\tt e}^2G_{\rm F}^{}}{4\sqrt2\,\pi^2}\, \overline s\gamma^\eta P_{L(R)}^{}d\, \overline\ell\gamma_\eta^{}\gamma_5^{}\ell \,, &
{\cal O}_{P(P')}^\ell & = \frac{\lambda_t^{}{\tt e}^2G_{\rm F}^{}m_s^{}}{4\sqrt2\,\pi^2}\, \overline sP_{R(L)}^{}d\, \overline\ell\gamma_5^{}\ell \,. 
\end{align} 
Thus, ${\cal H}_{\rm eff}^{\tt NP}$ causes changes to the amplitude component from short-distance SM physics, and the imaginary parts of the products $\lambda_tC_r$ constitute sources of $CP$ violation from~NP.
The combinations \,$C_{r\pm}\equiv\big(C_r\pm C_{r^\prime}\big)$,\, $r=7,9,10,S,P$,\, frequently show up hereafter.  

After the appropriate \,$\Sigma^+ {\,\to\,} p$\, matrix-elements of the quark bilinears in eqs.\,\,(\ref{O7}) and (\ref{Oj}) are applied, the SM and NP contributions can be put together in eq.\,(\ref{MS2pll}).
This has been implemented in ref.\,\cite{Roy:2024hqg}, and the outcomes are
\begin{align} \label{A...K} & 
\tilde{\textsc a} \,=\, \frac{G_{\rm F}^{}}{q^2} \bigg[ 4\alpha_{\tt 0}^{}\pi\, a - \frac{\alpha_{\tt Z}^{} c_\sigma^{} m_s^{}}{2\sqrt2\,\pi} \big(\lambda_u^* C_{7\gamma}^{} - 2 \lambda_t^*C_{7+}^*\big) \bigg] \,,
\nonumber \\ & 
\tilde{\textsc b} \,=\, \frac{G_{\rm F}^{}}{q^2} \bigg[ 4\alpha_{\tt 0}^{}\pi\, b - \frac{\alpha_{\tt Z}^{} c_\sigma^{} m_s^{}}{2\sqrt2\,\pi} \big(\lambda_u^* C_{7\gamma}^{} - 2 \lambda_t^* C_{7-}^*\big) \bigg] \,,
\nonumber \\ & 
\tilde{\textsc c} \,=\, G_{\rm F}^{} \Bigg[ 4\alpha_{\tt 0}^{}\pi\, c + \frac{\lambda_u^*z_{7V}^{}-\lambda_t^*y_{7V}^{}}{\sqrt2} -  \frac{\alpha_{\tt Z}^{} \lambda_t^* C_{9+}^{\ell*}}{2\sqrt2\,\pi} \Bigg] \,,
\nonumber \\ &  
\tilde{\textsc d} \,=\, G_{\rm F}^{} \Bigg[ 4\alpha_{\tt 0}^{}\pi\, d + \Bigg( \frac{\lambda_u^*z_{7V}^{}-\lambda_t^*y_{7V}^{}}{\sqrt2} - \frac{\alpha_{\tt Z}^{} \lambda_t^*{} C_{9-}^{\ell*}}{2\sqrt2\,\pi} \Bigg) (D{-}F) \Bigg] \,,  
\\
\tilde{\textsc e} =~ & \frac{\lambda_t^* G_{\rm F}^{}}{\sqrt2} \bigg( y_{7A}^{} + \frac{\alpha_{\tt Z}^{}}{2\pi}\, C_{10+}^{\ell*} \bigg) \,, \hspace{3em}
\tilde{\textsc f} = \frac{\lambda_t^* G_{\rm F}^{}}{\sqrt2} \bigg( y_{7A}^{} + \frac{\alpha_{\tt Z}^{}}{2\pi}\, C_{10-}^{\ell*} \bigg) (D{-}F) \,,  
\nonumber \\
\tilde{\textsc g} =~ & \frac{\alpha_{\tt Z}^{}G_{\rm F}^{} \lambda_t^* C_{\!S+}^{\ell*}}{2\sqrt2\,\pi} {\tt M}_-^{} \,, ~~~ \tilde{\textsc h} = \frac{\alpha_{\tt Z}^{}G_{\rm F}^{} \lambda_t^* C_{\!S-}^{\ell*} m_{\!K^0}^2 {\tt M}_+^{}}{2\sqrt2\,\pi~\big(m_{K^0}^2{-}q^2\big)} (D{-}F) \,, ~~~ 
\tilde{\textsc j} = \frac{-\alpha_{\tt Z}^{}G_{\rm F}^{} \lambda_t^*C_{\!P+}^{\ell*}}{2\sqrt2\,\pi} {\tt M}_-^{} \,, 
\nonumber \\ \nonumber
\tilde{\textsc k} =~ & \frac{\lambda_t^* G_{\rm F}^{}}{\sqrt2}\! \bigg[ 2 y_{7\!A}^{} m_\ell^{} + \frac{\alpha_{\tt Z}^{}}{\pi}\! \Big( C_{\!10-}^{\ell*} m_\ell^{} {+} C_{\!P-}^{\ell*} m_{\!K^0}^2\! \Big)\! \bigg]\! \frac{(D{-}F)\tt M_+^{}}{q^2{-}m_{\!K^0}^2} \,, ~~~~~~~ 
{\tt M}_\pm^{} = m_{\Sigma^+}^{} \pm m_p^{} \,, 
\end{align}  
where \,$\alpha_{\tt Z}^{}\equiv{\tt e}^2(m_Z)/(4\pi)=1/127.93$\, \cite{ParticleDataGroup:2024cfk}, we have neglected $m_d$ compared to $m_s$, and $c_\sigma^{}$ is a constant defined by \,$\langle p|\overline d\sigma^{\kappa\eta}\big(1,\gamma_5^{}\big)s|\Sigma^+\rangle
\varepsilon_\kappa^*q_\eta^{} = c_\sigma^{} \bar u_p^{} \sigma^{\kappa\eta} \big(1,\gamma_5^{}\big)u_\Sigma^{}\,
\varepsilon_\kappa^*q_\eta^{}$,\, with $\varepsilon$ being the photon-polarization four-vector.

From the resulting $\cal M$ and its antiparticle counterpart, the differential rate $d\Gamma/dq^2$ of \,$\Sigma^+ {\,\to\,} p\ell^+\ell^-$\, and that of \,$\overline\Sigma{}^- {\,\to\,} \overline p\ell^+\ell^-$\, can be calculated.
If the underlying interactions do not preserve $CP$ symmetry, the two rates could be unequal.
Consequently, a~quantity that can indicate $CP$ violation is
\begin{align} \label{dG'}
\delta\Gamma_\ell'\, & \equiv\, \frac{d\Gamma\big(\Sigma^+ {\,\to\,} p\ell^+\ell^-\big)}{dq^2} - \frac{d\Gamma\big(\overline\Sigma{}^- {\,\to\,} \overline p\ell^+\ell^-\big)}{dq^2}
\nonumber \\ & =\, \frac{\big(3\beta\!-\!\beta^3\big) \overline\lambda\raisebox{2pt}{$^{\frac{1}{2}}$} \alpha_{\tt O}^{} G_{\rm F}^2}{8\sqrt2\,\pi^2 m_{\Sigma^+}^3} \! \begin{array}[t]{l} \displaystyle \bigg\{ {\rm Im} \bigg[ {\tt M}_+^{} a + {\tt\hat m_+^2\,} \frac{c}{3} + q^2 c \bigg] {\rm\,Im} \bigg[ \lambda_t^{} y_{7V}^{} + \frac{\alpha_{\tt Z} \lambda_t}{2\pi} C_{9+}^\ell \bigg] \tt\,\hat m_-^2
\\ \displaystyle ~{\rm+\;Im} \bigg[ {\tt M}_-^{} b - {\tt\hat m_-^2\,} \frac{d}{3} - q^2 d \bigg] {\rm\,Im} \bigg[ \lambda_t^{} y_{7V}^{} + \frac{\alpha_{\tt Z} \lambda_t}{2\pi} C_{9-}^\ell \bigg] (F{-}D) {\tt\,\hat m_+^2} \bigg\} \end{array} ~~~ 
\nonumber \\ & ~~~~ +\, \frac{ \big(\beta^3\!-\!3\beta\big)
\overline\lambda\raisebox{2pt}{$^{\frac{1}{2}}$} \alpha_{\tt 0}^{} \alpha_{\tt Z}^{}G_{\rm F}^2 c_\sigma^{}m_s^{}}
{8\sqrt2\,\pi^3 m_{\Sigma^+}^3} \! \begin{array}[t]{l} \displaystyle {\rm\bigg\{Im} \bigg[ \overline\lambda\, \frac{2 a}{3q^2} + \big(a+{\tt M}_+^{}c\big)\, {\tt\hat m_-^2} \bigg] {\rm\,Im\,} {\tt\widetilde C_{7+}^{}}
\\ \displaystyle ~{\rm+\;Im} \bigg[ \overline\lambda\, \frac{2 b}{3q^2} + \big(b-{\tt M}_-^{}d\big) {\tt\,\hat m_+^2} \bigg] {\rm\,Im\,} {\tt\widetilde C_{7-}^{}} \bigg\} \,, \end{array} 
\end{align}  
where 
\begin{align} \label{tC7+-}
\beta = \sqrt{1-\frac{4m_\ell^2}{q^2}} \,, & & \overline\lambda = {\tt\hat m_-^2\hat m_+^2} \,, & & {\tt\hat m_\pm^2} = {\tt M_\pm^2} - q^2 \,, & & {\tt\widetilde C_{7\pm}^{}} = \lambda_t^{} C_{7\pm}^{} - \frac{\lambda_u}{2} C_{7\gamma}^* \,.
\end{align}

From $\delta\Gamma_\ell'$, we can devise the differential $CP$-asymmetries for \,$\ell=\mu,e$\, 
\begin{align} \label{Deltal'}
\hat\Delta_\mu' \,\equiv\, \frac{d\hat\Delta_\mu}{dq^2} \,=\, \frac{\tau_{\Sigma^+}^{}\, \delta\Gamma_\mu'}{2\,{\cal B}(\Sigma^+ {\to\,} p\mu^+\mu^-)_{\tt exp}^{}} \,, & & \hat\Delta_e' \,\equiv\, \frac{d\hat\Delta_e}{dq^2} \,=\, \frac{\tau_{\Sigma^+}^{}\, \delta\Gamma_e'}{\mathscr B_{ee}^{\tt SM}+\overline{\mathscr B}{}_{ee}^{\tt SM}} \,, 
\end{align}  
where $\tau_{\Sigma^+}$ is the $\Sigma^+$ lifetime, 
\begin{align} 
\mathscr B_{\ell\ell}^{\tt SM} & \,\equiv\, {\cal B}\big(\Sigma^+ {\to\,} p\ell^+\ell^-\big){}_{\tt SM}^{} \,, & \overline{\mathscr B}{}_{\ell\ell}^{\tt SM} & \,\equiv\, {\cal B}\big(\overline\Sigma{}^- {\to\,} \overline p\ell^+\ell^-\big){}_{\tt SM}^{} \,. ~~~ ~~~
\end{align}  
For the normalization of $\hat\Delta_\mu'$ we have selected the empirical value in eq.\,(\ref{Bexp}), which has good precision and implicitly incorporates the $\Sigma^+$ and $\overline\Sigma{}^-$ processes \cite{LHCb:2025evf}, and so no theoretical error is introduced in the denominator of $\hat\Delta_\mu'$.  
In the case of $\hat\Delta_e'$, we can alternatively normalize it with respect to ${\cal B}(\Sigma^+ {\,\to\,} pe^+e^-)_{\tt exp}$ when the latter has become sufficiently well-determined in the future.

Numerically, we employ the long-distance form factors ($a,b,c,d$) corresponding\linebreak to the aforesaid ``relativistic solution 4'', as detailed in ref.\,\,\cite{Roy:2024hqg}.
Moreover, we adopt\linebreak$z_{7V} = (-0.05\pm0.02)\,\alpha_{\rm Z}$,\, $y_{7V} = (0.73\pm0.04)\,\alpha_{\rm Z}$,\, $y_{7A} = (-0.68\pm0.03)\,\alpha_{\rm Z}$\, \cite{Buchalla:1995vs,Isidori:2004rb,Neshatpour:2022fak},\linebreak$C_{7\gamma}=0.72-0.79\,i{\rm\,\,Im}\,\lambda_t$\, \cite{Roy:2024hqg,Nielsen:1995dp,Tandean:1999mg}, and \,$m_s=(123\pm1)$~MeV,\, all evaluated at the re-\linebreak normalization scale of 1\,\,GeV.
Additionally, we use the CKM parameters, which lead\linebreak to \,$\lambda_t = [3.23\pm0.14+(1.42\pm0.06)i]\times10^{-4}$,\, hadron and lepton masses, and $\Sigma^+$ lifetime\linebreak from ref.\,\cite{ParticleDataGroup:2024cfk}, \,$c_\sigma^{}=0.30\pm0.09$\, from quark-model computations \cite{Gilman:1978bg,Donoghue:2022wrw}, and \,$D = 0.81\pm0.01$  and \,$F = 0.47\pm0.01$\, from fitting to the data on semileptonic baryon decays \cite{ParticleDataGroup:2024cfk,Cabibbo:2003cu}.
Some other pertinent constants are collected in appendix~\ref{smrates}. 
The uncertainties of the different input parameters and data are reflected in the various branching-fraction and $CP$-asymmetry outcomes exhibited below.

Now, the \,$\ell=\mu$\, branching-fraction containing the SM part and the interference between its long-distance portion and the NP pieces has already been given in ref.\,\cite{Roy:2024hqg} in the $CP$-symmetric limit.
Supplementing that result with the $CP$-violating contributions and specializing to the relativistic solution~4, we  obtain for the $\Sigma^+$ and $\overline\Sigma{}^-$ channels as well as the \,$\ell=e$\, ones
\begin{align}
{\cal B}(\Sigma^+ {\to\,} p\mu^+\mu^-) & = \mathscr B_{\mu\mu}^{\tt SM} + \raisebox{3pt}{\scriptsize$\displaystyle\sum_{\mbox{\it\scriptsize r}}$} \Big[ \rho_r^\mu\, {\rm Re}\big(\lambda_t^{}C_r^\mu\big) + \eta_r^\mu\, {\rm Im}\big(\lambda_t^{}C_r^\mu\big) \Big] \!\times 10^{-8} + \cdots \,,
\nonumber \\
{\cal B}\big(\overline\Sigma{}^- {\to\,} \overline p\mu^+\mu^-\big) & = \overline{\mathscr B}{}_{\mu\mu}^{\tt SM} + \raisebox{3pt}{\scriptsize$\displaystyle
\sum_{\mbox{\it\scriptsize r}}$} \Big[ \rho_r^\mu\, {\rm Re}\big(\lambda_t^{}C_r^\mu\big) - \eta_r^\mu\, {\rm Im}\big(\lambda_t^{}C_r^\mu\big) \Big] \!\times 10^{-8} + \cdots \,, ~~~
\nonumber \\
{\cal B}(\Sigma^+ {\to\,} pe^+e^-) & = \mathscr B_{ee}^{\tt SM} + \raisebox{3pt}{\scriptsize$\displaystyle\sum_{\mbox{\it\scriptsize r}}$} \Big[ \rho_r^e\, {\rm Re}\big(\lambda_t^{}C_r^e\big) + \eta_r^e\, {\rm Im}\big(\lambda_t^{}C_r^e\big) \Big] \!\times 10^{-6} + \cdots \,,
\nonumber \\
{\cal B}\big(\overline\Sigma{}^- {\to\,} \overline pe^+e^-\big) & = \overline{\mathscr B}{}_{ee}^{\tt SM} + \raisebox{3pt}{\scriptsize$\displaystyle\sum_{\mbox{\it\scriptsize r}}$} \Big[ \rho_r^e\, {\rm Re}\big(\lambda_t^{}C_r^e\big) - \eta_r^e\, {\rm Im}\big(\lambda_t^{}C_r^e\big) \Big] \!\times 10^{-6} + \cdots \,, \label{eq:interf}
\end{align}  
where 
\begin{align} \label{smBll}
\mathscr B_{\mu\mu}^{\tt SM} & \,=\, (1.2\pm0.1) \times 10^{-8} \,=\, \overline{\mathscr B}{}_{\mu\mu}^{\tt SM} + (2.3\pm0.2) \times 10^{-12} \,, &
\nonumber \\ 
\mathscr B_{ee}^{\tt SM} & \,=\, (7.4\pm0.3) \times 10^{-6} \,=\, \overline{\mathscr B}{}_{ee}^{\tt SM} + (6.6\pm0.5) \times 10^{-11} \,,
\end{align}
the factors $\rho_r^\ell$ and $\eta_r^\ell$ for \,$\ell=\mu,e$\, and \,$r=7\pm,9\pm$\, have been gathered in table\,\,\ref{rho,eta}, and the ellipses represent terms with much smaller $\rho$s and $\eta$s (the latter including \pagebreak 
$CP$-even ones arising from interference with the SM short-distance amplitudes) and terms of second order in the~$C$s.
The $\mathscr B_{\mu\mu}^{\tt SM}$ and $\mathscr B_{ee}^{\tt SM}$ numbers in eq.\,(\ref{smBll}) are unchanged compared to the corresponding values reported in ref.\,\cite{Roy:2024hqg}, after the most recent input-data have been taken into account, as outlined in appendix \ref{smrates}.

\begin{table}[t]\centering\setlength{\tabcolsep}{1ex}\begin{tabular}{|c||c|c|c|c|}\hline 
$\ell\vphantom{|_|^|}$ & $\rho_{7+}^\ell$ & $\rho_{7-}^\ell$ & $\rho_{9+}^\ell$ & $\rho_{9-}^\ell$ \\ \hline
$\mu$ & $0.07\pm0.02$ & $-0.14\pm0.05$ & $-0.021\pm0.001$ & $-0.14\pm0.02$ \\
$e$ & $0.7\pm0.2$ & $-0.29\pm0.09$ & $-0.0098\pm0.0004$ & $-0.024\pm0.003$ \\\hline\end{tabular}\vspace{3pt}\\\begin{tabular}{|c||c|c|c|c|}\hline$\ell\vphantom{|_|^|}$ & $\eta_{7+}^\ell$ & $\eta_{7-}^\ell$ & $\eta_{9+}^\ell$ & $\eta_{9-}^\ell$ \\ \hline
$\mu$ & $-0.03\pm0.01$ & $0.10\pm0.03$ & $0.082\pm0.001$ & $0.099\pm0.004$ \\
$e$ & $-0.18\pm0.05$ & $0.11\pm0.03$ & $0.0371\pm0.0006$ & $0.0126\pm0.0006$ \\ \hline
\end{tabular}\caption{The $\rho_r^\ell$ and $\eta_r^\ell$ factors in eq.\,(\ref{eq:interf}) for \,$\ell=\mu,e$\, and \,$r=7\pm,9\pm$.\label{rho,eta}}
\end{table}

From eq.\,(\ref{eq:interf}), or more accurately from the integration of \,$d\hat\Delta_\ell/dq^2$\, in eq.\,(\ref{Deltal'}) over the whole kinematical interval \,$4m_\ell^2\le q^2\le \big(m\raisebox{-1pt}{$_{\Sigma^+}$}{-}\,m_p\big){}^2$\, for \,$\ell=\mu,e$,\, we then arrive at the rate asymmetries 
\begin{align} \label{Deltamu}
\hat\Delta_\mu^{}\, & =\, \frac{{\cal B}\big(\Sigma^+ {\,\to\,} p\mu^+\mu^-\big) - {\cal B}\big(\overline\Sigma{}^- {\,\to\,} \overline p\mu^+\mu^-\big)}{2\,{\cal B}(\Sigma^+ {\,\to\,} p\mu^+\mu^-)_{\tt exp}^{}}
\nonumber \\
& =\, \hat\Delta_\mu^{\tt SM} - (0.03\pm0.01) {\rm\,Im} \big(\lambda_t^{}C_{7+}^{}\big) + (0.09\pm0.03) {\rm\,Im} \big(\lambda_t^{}C_{7-}^{}\big) 
\nonumber \\ & ~~~ +\, (0.07\pm0.01) {\rm\,Im} \big(\lambda_t^{}C_{9+}^\mu\big) + (0.09\pm0.01) {\rm\,Im} \big(\lambda_t^{}C_{9-}^\mu\big) \,,
\nonumber \\ \raisebox{6ex}{}
\hat\Delta_e^{}\, & =\, \frac{{\cal B}\big(\Sigma^+ {\,\to\,} pe^+e^-\big) - {\cal B}\big(\overline\Sigma{}^- {\,\to\,} \overline pe^+e^-\big)}{\mathscr B_{ee}^{\tt SM}+\overline{\mathscr B}{}_{ee}^{\tt SM}}
\nonumber \\
& =\, \hat\Delta_e^{\tt SM} - (0.024\pm0.007) {\rm\,Im} \big(\lambda_t^{}C_{7+}^{}\big) + (0.015\pm0.005) {\rm\,Im} \big(\lambda_t^{}C_{7-}^{}\big) 
\nonumber \\ & ~~~ +\, (0.0050\pm0.0002) {\rm\,Im} \big(\lambda_t^{}C_{9+}^e\big) + (0.0017\pm0.0001) {\rm\,Im} \big(\lambda_t^{}C_{9-}^e\big) \,, ~~~
\end{align}
where 
\begin{align} \label{sm-Deltal}
\hat\Delta_\mu^{\tt SM} & \,=\, (1.1\pm0.2)\times10^{-4} \,, & \hat\Delta_e^{\tt SM} & \,=\, (4.5\pm0.4)\times10^{-6} ~~~ ~~
\end{align}
are supplied by the SM. 
Clearly, among the coefficients possibly induced by NP, only Im$(\lambda_tC_{7\pm})$ and Im$\big(\lambda_t^{}C_{9\pm}^\ell\big)$ can be probed by $\hat\Delta_\ell$.

Turning to the weak radiative process \,$\Sigma^+ {\,\to\,} p\gamma$,\, which emits an onshell photon, we write the amplitude for this decay and its rate, respectively, as
\begin{align}
\mathscr M_\gamma^{} & = i{\tt e} G_{\rm F}^{}\, \bar u_p^{} \big({\tt A}_\gamma\!+\!\gamma_5^{}{\tt B}_\gamma\big) \sigma^{\kappa\eta} u_\Sigma^{}\, \varepsilon_\kappa^* q_\eta^{} \,, &  
\Gamma_\gamma^{} & = \frac{{\tt e}^2G_{\rm F}^2}{8\pi} \frac{\big(m_{\Sigma^+}^2\!-\!m_p^2\big)\raisebox{1pt}{$^3$}}{m_{\Sigma^+}^3} \big(|{\tt A}_\gamma|^2\!+\!|{\tt B}_\gamma|^2\big) \,,
\end{align}
where ${\tt A}_\gamma$ and ${\tt B}_\gamma$ comprise long- and short-distance components, the latter involving the SM coefficient $C_{7\gamma}$ and the NP ones $C_{7\pm}$.
Explicitly
\begin{align} \label{AgBg}
{\tt A}_\gamma & \,=\, a(0) + \frac{c_\sigma^{} {\tt \widetilde C_{7+}^*} m_s^{}}{4\sqrt2\,\pi^2} \,, & {\tt B}_\gamma & \,=\, b(0) + \frac{c_\sigma^{} {\tt \widetilde C_{7-}^*} m_s^{}}{4\sqrt2\,\pi^2} \,, ~~~ ~~~~ 
\end{align}
where $a(0)$ and $b(0)$ are the \,$q^2=0$\, values of $a$ and $b$, the weak phases in them having been ignored, and \,${\tt\widetilde C_{7\pm}^{}} = \lambda_t^{} C_{7\pm}^{} - \lambda_u^{} C_{7\gamma}^*/2$\, as defined in eq.\,(\ref{tC7+-}).
The antihyperon decay \,$\overline\Sigma{}^- {\,\to\,} \overline p\gamma$\, has analogous formulas but with ${\tt A}_\gamma$ and ${\tt B}_\gamma$ replaced by
\begin{align} \label{bAgbBg}
\overline{\tt A}_\gamma & \,=\, -a(0) - \frac{c_\sigma^{} {\tt \widetilde C_{7+}^{}} m_s^{}}{4\sqrt2\,\pi^2} \,, & \overline{\tt B}_\gamma & \,=\, b(0) + \frac{c_\sigma^{} {\tt \widetilde C_{7-}^{}} m_s^{}}{4\sqrt2\,\pi^2} \,, ~~~ ~~~
\end{align}
as \,$\lambda_u(\simeq0.2192)$\, is real in the standard parameterization.

One can therefore construct the $CP$-violating observables 
\begin{align} \label{A,Delta}
\hat\Delta_\gamma^{}\, & \equiv\, \frac{{\cal B}(\Sigma^+ {\,\to\,} p\gamma) - {\cal B}\big(\overline\Sigma{}^-  {\,\to\,} \overline p\gamma\big)}{2\,{\cal B}(\Sigma^+ {\,\to\,} p\gamma)_{\tt exp}^{}} \,, & \hat A_\gamma^{} & \,\equiv\, \frac{\alpha_\gamma+\overline\alpha_\gamma}{2\,\alpha_{\Sigma^+\to p\gamma}^{\tt exp}} \,, ~~~ 
\end{align}
where $\hat A_\gamma$ contains the decay asymmetry parameters 
\begin{align} 
\alpha_\gamma^{} \,=\, \frac{2\, {\rm Re}\big( {\tt A}_\gamma^* {\tt B}_\gamma^{}\big)}{|{\tt A}_\gamma|^2 + |{\tt B}_\gamma|^2} \,, & & \overline\alpha_\gamma^{} \,=\, \frac{2\, {\rm Re}\big( \overline{\tt A}{}_\gamma^* \overline{\tt B}_\gamma^{}\big)}{|\overline{\tt A}_\gamma|^2 + |\overline{\tt B}_\gamma|^2} \,, & ~~~ ~~~~
\end{align}
and the empirical $\alpha_{\Sigma^+\to p\gamma}^{\tt exp}$,  
which incorporates the $\Sigma^+$ and $\overline\Sigma{}^-$ results \cite{BESIII:2023fhs}. 
The asymmetries in eq.\,(\ref{A,Delta}) are of interest here because they offer another access to Im$(\lambda_tC_{7\pm})$ and were probed for the first time by BESIII not long ago \cite{BESIII:2023fhs}. 
Expanding eq.\,(\ref{A,Delta}) in terms of imaginary parts, we have 
\begin{align} \label{Delta-g}
\hat\Delta_\gamma^{} & \,=\, \frac{\sqrt{\alpha_{\tt0}^{}\alpha_{\tt Z}^{}} \big(m_{\Sigma^+}^2 \!-\! m_p^2\big)\raisebox{1pt}{$^3$} \tau_{\Sigma^+}^{} G_{\rm F}^2 c_\sigma^{} m_s^{}}{4\sqrt2\,\pi^2~ {\cal B}(\Sigma^+ {\to\,} p\gamma)_{\tt exp}^{}\, m_{\Sigma^+}^3} \Big[ {\rm-Im}\hspace{1pt}a(0){\rm\,Im}{\tt\,\widetilde C_{7+}^{}} - {\rm Im}\hspace{1pt}b(0){\rm\,Im}{\tt\,\widetilde C_{7-}^{}} \Big] \,,
\end{align}
which is exact, and 
\begin{equation} \label{A-g} \!\!\!\begin{array}[b]{rl} \displaystyle 
\hat A_\gamma^{} \,=\, \frac{-\big(m_{\Sigma^+}^2 -m_p^2\big)\raisebox{1pt}{$^3$}\, \tau_{\Sigma^+}^{} \, G_{\rm F}^2}{\alpha_{\Sigma^+\to p\gamma}^{\tt exp} {\cal B}(\Sigma^+ {\to} p\gamma)_{\tt exp}^{} m_{\Sigma^+}^3} \!\! & \Bigg\{ \displaystyle c_\sigma^{} m_s^{} \sqrt{\alpha_{\tt0}^{}\alpha_{\tt Z}^{}}\, \frac{{\rm Im}\hspace{1pt}a(0){\rm\,Im}\hspace{1pt}{\tt\widetilde C_{7-}} {+} {\rm\,Im}\hspace{1pt}b(0){\rm\,Im}\hspace{1pt}\tt\widetilde C_{7+}}{4\sqrt2\,\pi^2} \\ & ~~+~ \alpha_{\tt0}^{}\, \hat\Delta_\gamma^{} {\,\rm Re}\big[a^*(0)\,b(0)\big] \Bigg\} \,, \end{array} 
\end{equation}
which is valid to first order in the $C$s.
With the $\alpha_{\Sigma^+\to p\gamma}^{\tt exp}$, ${\cal B}(\Sigma^+ {\to} p\gamma)_{\tt exp}$, $a(0)$, and $b(0)$ values enumerated in appendix \ref{smrates}, we then get 
\begin{align} \label{Ag,Dg}
\hat\Delta_\gamma^{} & \,=\, \hat\Delta_\gamma^{\tt SM} - (0.022\pm0.007) {\rm\,Im} (\lambda_tC_{7+}) + (0.013\pm0.004) {\rm\,Im} (\lambda_tC_{7-}) \,,
\nonumber \\
\hat A_\gamma^{} & \,=\, \hat A_\gamma^{\tt SM} + (0.003\pm0.002) {\rm\,Im} (\lambda_tC_{7+}) + (0.019\pm0.006) {\rm\,Im} (\lambda_tC_{7-}) \,, ~
\end{align}
where
\begin{align} \label{sm-Deltag}
\hat\Delta_\gamma^{\tt SM} & \,=\, (1.1\pm0.3) \times10^{-7} \,, &
\hat A_\gamma^{\tt SM} & \,=\, (-2.7\pm0.9) \times10^{-7} ~~~ 
\end{align}
originate from the SM.\footnote{Early  attempts to evaluate $\hat \Delta_\gamma^{\tt SM}$ and $\hat A_\gamma^{\tt SM}$ were made in refs.\,\cite{Chen:1995vw,Li:1993hp}.\medskip}

\vspace{2ex}

\pagebreak

It is noteworthy that the real parts of the long-distance form factors $(a,b,c,d)$ do not enter eq.\,\,(\ref{dG'}), (\ref{Delta-g}), or~(\ref{A-g}) singly. 
This means that the $CP$-violating observables being investigated $\big(\hat\Delta_\mu,\hat\Delta_e,\hat\Delta_\gamma$, and~$\hat A_\gamma\big)$ are unaffected by the ambiguity in the values of Re\,$a$ and Re\,$b$ extracted from \,$\Sigma^+ {\to\,} p\gamma$\, data\footnote{The product \,Re$\hspace{1pt}a\,$Re$\hspace{1pt}b$\, enters Re$[a^*b]$ in eq.\,(\ref{A-g}) but is free from this ambiguity, as can easily be verified from, e.g., Table I in ref.\,\cite{Roy:2024hqg}.\medskip} or by the uncertainty in the estimation of Re\,$c$ and Re\,$d$, which are both aspects of the theoretical treatment elaborated in ref.\,\cite{He:2005yn}.
Their imaginary parts, however, still depend on whether they are computed using the relativistic or heavy-baryon formulation of chiral perturbation theory~\cite{He:2005yn,Roy:2024hqg}, but the former (namely with the solution 4) has turned out to lead to predictions favored by experiment,\footnote{More specifically, the predictions of heavy-baryon chiral perturbation theory for ${\cal B}(\Sigma^+ {\to\,} p\mu^+\mu^-)$ are disfavored by at least 3.1$\sigma$ compared to the preferred prediction of relativistic chiral perturbation\linebreak theory \cite{LHCb:2025evf,Roy:2024hqg}.
In regard to the other relativistic solutions (1-3, as discussed in ref.\,\cite{Roy:2024hqg}), they are rejected at over 6.1$\sigma$ by the LHCb measurement \cite{LHCb:2025evf}.\medskip}
as mentioned in section \ref{intro}.

\section{New physics contributions\label{np}}

\subsection{Model-independent constraints\label{modindep}}

To see the extents to which $\hat\Delta_\ell$, $\hat\Delta_\gamma$, and $\hat A_\gamma$ could be enhanced over their SM expectations by the NP potentially encoded in $C_{7\pm}^{}$ and $C_{9\pm}^\ell$, we may suppose model-independently that only one of them is nonzero at a~time in order to infer its maximum size as permitted by the relevant experimental restrictions.
Under this assumption, the strictest bounds on Im$(\lambda_tC_{7+})$ and Im$(\lambda_tC_{7-})$ can be extracted from the measurements \cite{ParticleDataGroup:2024cfk} of \,$K
\!_L \to \pi^0e^+e^-$  and \,$K^+ \to \pi^+\pi^0\gamma$,\, respectively \cite{Mertens:2011ts}.
Accordingly, employing the latest lattice assessment of the kaon-to-pion matrix element of the $\sigma$ tensor,\footnote{We have \,$\langle\pi^0|\overline s\sigma^{\kappa\tau}d|K^0\rangle = -\langle\pi^0|\overline d\sigma^{\kappa\tau}s|\,\overline{\!K}{}^0\rangle = i\sqrt2\big(P_\pi^\kappa P_K^\tau-P_K^\kappa P_\pi^\tau\big)f_T^{}(0)/\big[\big(1-q^2s_T^{}\big)(m_\pi+m_K)\big]$,  with \,$f_T^{}(0) = 0.47\pm0.02$\, and \,$s_T^{}=(1.10\pm0.14)\rm/GeV^2$\, from ref.\,\cite{Baum:2011rm}, the $f_T^{}(0)$ value having been translated from that at a renormalization scale of 2\,GeV to that at 1\,GeV using the pertinent formula from ref.\,\cite{Becirevic:2000zi}.\medskip} we find
\begin{align} \label{indepImC7} 
{-}0.032 < {\rm Im}(\lambda_tC_{7+}) < {+}0.017 \,, & & {-}0.19 < {\rm Im}(\lambda_tC_{7-}) < 0.41 \,. & ~~~ ~~~~
\end{align}

In a similar way, the allowed ranges of Im$\big(\lambda_t^{}C_{9+}^\ell\big)$ can be derived from the data\linebreak ${\cal B}\big(K\!_L^{} \to \pi^0\mu^+\mu^-\big){}_{\tt exp} < 3.8\times10^{-10}$\, and \,${\cal B}\big(K\!_L^{} \to \pi^0 e^+e^-\big){}_{\tt exp} < 2.8\times10^{-10}$\, \cite{AlaviHarati:2000hs,AlaviHarati:2003mr}, both at 90\% confidence level (CL).
Thus, we apply their theoretical expressions as given in refs.\,\,\cite{Roy:2024hqg,Mescia:2006jd,DAmbrosio:2022kvb} to obtain, respectively,
\begin{align} \label{ImC9+-ranges}
{-}0.018 < {\rm Im}\big(\lambda_t^{}C_{9+}^\mu\big) < {+}0.014 \,, & & {-}0.0084 < {\rm Im}\big(\lambda_t^{}C_{9+}^e\big) < {+}0.0046 \,. & ~~~  
\end{align}

As for Im$\big(\lambda_t^{}C_{9-}^\ell\big)$, the most consequential restraints to date have turned out to\linebreak come from the hyperon sector, namely the deviations of the measurements in eq.\,(\ref{Bexp})\linebreak and \,\mbox{${\cal B}(\Sigma^+ {\to\,} pe^+e^-)_{\tt exp} = (7.7\pm4.6) \times 10^{-6\,}$ \cite{He:2005yn,Ang:1969hg}} from the respective SM predictions\linebreak cited in eq.\,(\ref{smBll}).
This implies the relatively much weaker limits\hspace{1pt}\footnote{Since the LHCb analysis yielding eq.\,(\ref{Bexp}) implicitly covered both $\Sigma^+$ and $\overline\Sigma{}^-$ processes~\cite{LHCb:2025evf}, to constrain Im$\hspace{1pt}\widetilde{\textsf c}{}_{9-}^\mu$, where \,$\widetilde{\textsf c}{}_{9-}^\ell \!\equiv\! \lambda_t^{}C_{9-}^\ell$,\, we first equate $2{\cal B}(\Sigma^+ {\to\hspace{1pt}} p\mu^+\mu^-)_{\tt exp}$ to the theoretical formula ${\cal B}(\Sigma^+ {\to\hspace{1pt}} p\mu^+\mu^-) + {\cal B}(\overline\Sigma{}^- {\to\hspace{1pt}} \overline p\mu^+\mu^-) = \mathscr B_{\mu\mu}^{\tt SM} + \overline{\mathscr B}{}_{\mu\mu}^{\tt SM} + 10^{-13} (5.0\pm0.5) {\rm\,Im}\hspace{1pt} \widetilde{\textsf c}{}_{9-}^\mu + 10^{-10} (3.9\pm0.3) \big({\rm Im}\hspace{1pt}\widetilde{\textsf c}{}_{9-}^\mu\big){}^2$, upon fixing \,Re$\hspace{1pt}\widetilde{\textsf c}{}_{9-}^\mu = C_{r\neq9-} =0$,\, where the part linear in Im$\hspace{1pt}\widetilde{\textsf c}{}_{9-}^\mu$ is $CP$ even and arises from interference with the SM short-distance components. 
To constrain Im$\hspace{1pt}\widetilde{\textsf c}{}_{9-}^e$, given that the quoted ${\cal B}(\Sigma^+ {\to\hspace{1pt}} pe^+e^-)_{\tt exp}$ above was determined solely from $\Sigma^+$ events~\cite{Ang:1969hg}, we set it equal to the formula
\,${\cal B}(\Sigma^+ {\to\hspace{1pt}} pe^+e^-) = \mathscr B_{ee}^{\tt SM} + 10^{-8} (1.3\pm0.1) {\rm\,Im}\hspace{1pt}\widetilde{\textsf c}{}_{9-}^e + 10^{-9} (1.7\pm0.1) \big({\rm Im}\hspace{1pt}\widetilde{\textsf c}{}_{9-}^e\big){}^2$, after removing Re$\hspace{1pt}\widetilde{\textsf c}{}_{9-}^e$ and $C_{r\neq9-}$. 
We subsequently require that Im$\hspace{1pt}\widetilde{\textsf c}{}_{9-}^\mu$ and Im$\hspace{1pt}\widetilde{\textsf c}{}_{9-}^e$ satisfy the 90\%-CL intervals of \,$2{\cal B}(\Sigma^+ {\to\hspace{1pt}} p\mu^+\mu^-)_{\tt exp} {\,-\,} \mathscr B_{\mu\mu}^{\tt SM} {\,-\,} \overline{\mathscr B}{}_{\mu\mu}^{\tt SM}$\, and \,${\cal B}(\Sigma^+ {\to\hspace{1pt}} pe^+e^-)_{\tt exp} {\,-\,} \mathscr B_{ee}^{\tt SM}$,\, respectively. 
This leads us finally to eq.\,(\ref{ImC9--ranges}).}
\begin{align} \label{ImC9--ranges} 
{-}3.5 < {\rm Im}\big(\lambda_t^{}C_{9-}^\mu\big) < 3.5 \,, & & {-}77 < {\rm Im}\big(\lambda_t^{}C_{9-}^e\big) < {+}68 \,. & ~~~ ~~~~~
\end{align}
Though $C_{9-}^\ell$ impacts \,$K {\to\,} \pi\pi\ell^+\ell^-$\, as well, the bounds deduced from their current experimental rates are comparatively less important.
In table~\ref{wccons} we have summarized the permitted values in eqs.\,(\ref{indepImC7})-(\ref{ImC9--ranges}).

\begin{table}[t]\centering\setlength{\tabcolsep}{1ex}\begin{tabular}{|c|c|c|}\hline WC & 90\%-CL range & Process$\vphantom{|_|^|}$ \\ \hline
Im$\big(\lambda_t^{}C_{7+}^{}\big)$ & $(-0.032,\, 0.017)$ & $K_L\to\pi^0e^+e^-$ \\
Im$\big(\lambda_t^{}C_{7-}^{}\big)$ & $(-0.19,\, 0.41)$ & $K^+\to \pi^+\pi^0\gamma$ \\
Im$\big(\lambda_t^{}C_{9+}^\mu\big)$ & $(-0.018,\, 0.014)$ & ~$K_L\to \pi^0\mu^+\mu^-$~ \\
Im$\big(\lambda_t^{}C_{9+}^e\big)$ & ~$(-0.0084,\, 0.0046)$~ & $K_L\to \pi^0e^+e^-$ \\
Im$\big(\lambda_t^{}C_{9-}^\mu\big)$ & $(-3.5,\, 3.5)$ & $\Sigma^+\to p\mu^+\mu^-$ \\
Im$\big(\lambda_t^{}C_{9-}^e\big)$ & $(-77,\, 68)$ & $\Sigma^+\to pe^+e^-$ \\ \hline
\end{tabular}\caption{The allowed (90\% CL) ranges of the imaginary parts of the LEFT Wilson coefficients (WCs) from possible NP that influences the $CP$-violating rate asymmetry of \,$\Sigma^+ {\,\to\,} p\ell^+\ell^-$\, if merely one of them contributes at a time and a model-independent approach is adopted.
The process imposing the most significant restriction in each case is also listed.\label{wccons}}
\end{table}

\begin{figure}[t]\bigskip\centering\hspace{-1ex}\includegraphics[width=74mm]{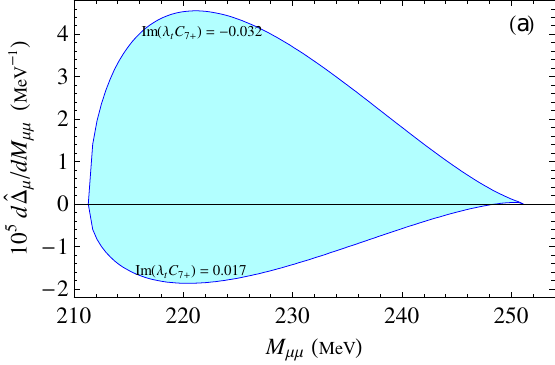} ~~ \includegraphics[width=74mm]{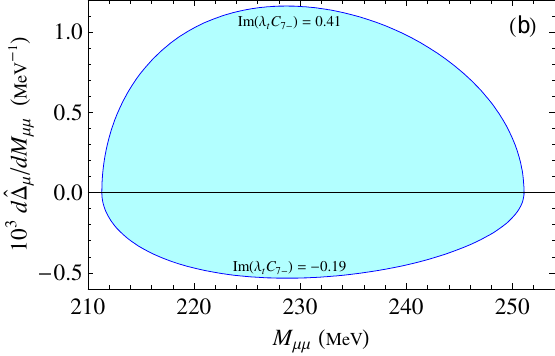}\vspace{5pt}\\
\hspace{-1ex}\includegraphics[width=74mm]{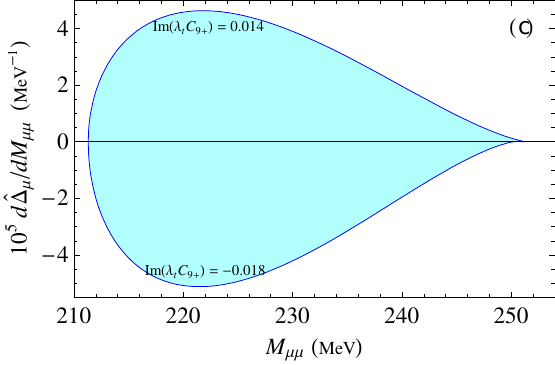} ~~ \includegraphics[width=74mm]{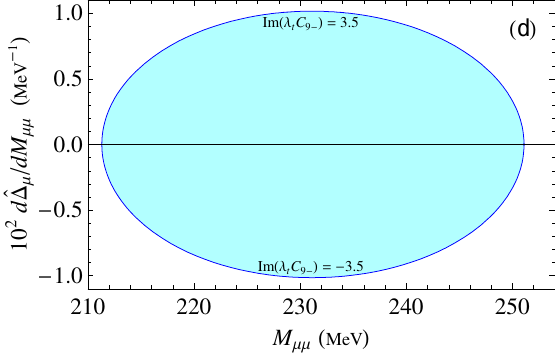}\vspace{-5pt}\caption{The range of the differential $CP$-asymmetry $d\hat\Delta_\mu/dM_{\mu\mu}$ of \,$\Sigma^+ {\,\to\,} p\mu^+\mu^-$ (cyan)  versus the dimuon invariant mass $M_{\mu\mu}$ if new physics induces (a)\,\,Im$(\lambda_tC_{7+})$, (b)\,\,Im$(\lambda_tC_{7-})$, (c)\,\,Im$\big(\lambda_t^{}C_{9+}^\mu\big)$, or (d)\,\,Im$\big(\lambda_t^{}C_{9-}^\mu\big)$ alone, with its interval given in table~\ref{wccons}.
The Im$(\lambda_tC_r)$ values corresponding to the upper and lower boundaries of each cyan area are also shown.} \label{Deltamu'_vs_Mmumu}
\end{figure}

\begin{figure}[t]\centering\hspace{-1ex}\includegraphics[width=74mm]{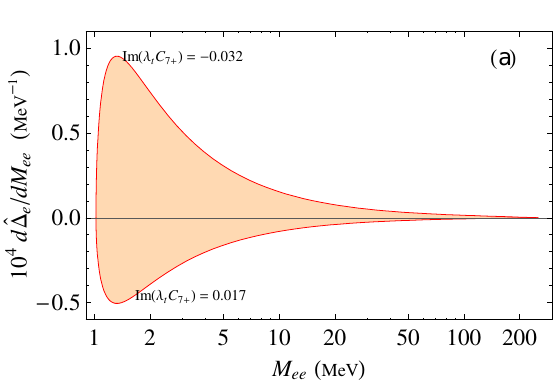} ~~ \includegraphics[width=74mm]{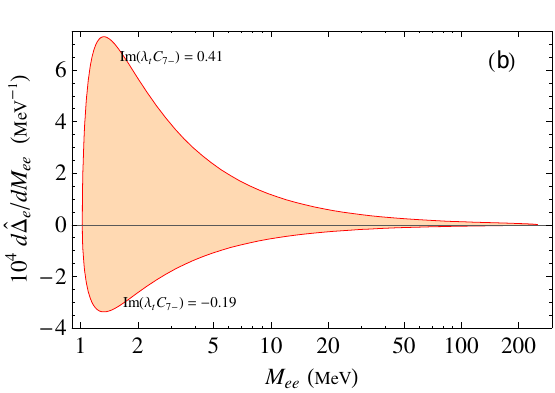}\vspace{-3pt}\\\hspace{-1ex}\includegraphics[width=74mm]{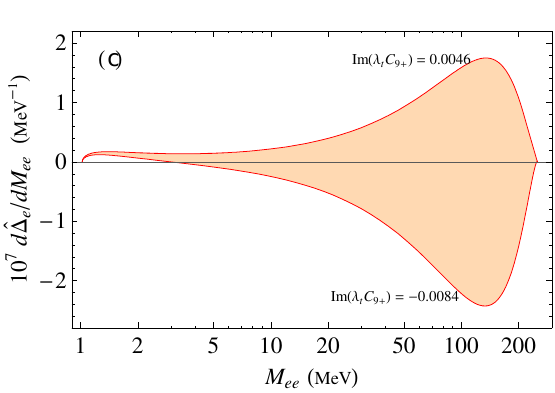} ~~ \includegraphics[width=74mm]{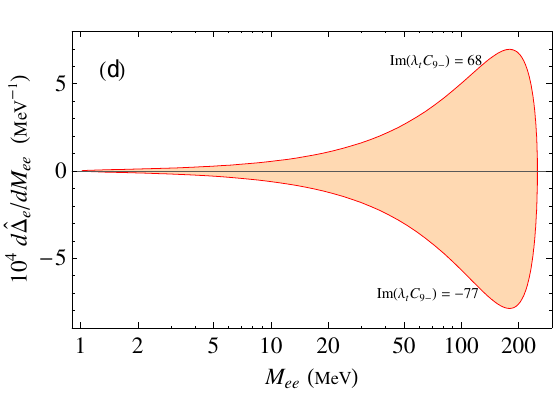}\vspace{-1ex}\caption{The range of the differential $CP$-asymmetry $d\hat\Delta_e/dM_{ee}$ of \,$\Sigma^+ {\,\to\,} pe^+e^-$ (orange) versus the dielectron invariant mass $M_{ee}$ if new physics induces (a)~Im$(\lambda_tC_{7+})$, (b)~Im$(\lambda_tC_{7-})$, (c)~Im$\big(\lambda_t^{}C_{9+}^e\big)$, or (d)~Im$\big(\lambda_t^{}C_{9-}^e\big)$ alone, with its interval given in table \ref{wccons}.
The Im$(\lambda_tC_r)$ values corresponding to the upper and lower boundaries of each orange area are also shown.} \label{Deltae'_vs_Mee} 
\end{figure}

To provide some visual illustrations of the effects of \,Im$(\lambda_tC_r)$, \,$r\!=\!7\pm,9\pm$,\, being present one at a time, we have plotted the distributions of \,$d\hat\Delta_\ell^{}/dM_{\ell\ell}^{} = 2M_{\ell\ell}^{}\,\hat\Delta_\ell'$\, in relation to the dilepton invariant mass \,$M_{\ell\ell}^{}=\big(q^2\big)\raisebox{1pt}{$^{1/2}$}$,\, as dictated by eq.\,(\ref{Deltal'}), for \,$\ell=\mu,e$\, in figures \ref{Deltamu'_vs_Mmumu} and \ref{Deltae'_vs_Mee}, respectively.
From figure \ref{Deltae'_vs_Mee}, we learn that the Im$(\lambda_tC_{7\pm})$ contributions to $d\hat\Delta_e/dM_{ee}$ are strongly peaked at low $M_{ee}$, which is attributable to the $1/q^2$ terms in~eq.\,(\ref{dG'}). 
Needless to say, both $d\hat\Delta_\ell/dM_{\ell\ell}$ and its integrated form, $\hat\Delta_\ell$, are highly desirable to pursue experimentally, as the outcomes could reveal complementary information on the underlying $CP$-violating interactions. 

Putting together eqs.\,\,(\ref{indepImC7})-(\ref{ImC9--ranges}) with eqs.\,\,(\ref{Deltamu}) and (\ref{Ag,Dg}) including the SM portions, and supposing that only one of the NP coefficients is existent in each instance, we arrive at the following ranges of the various $CP$-asymmetries being studied:
\begin{align} \label{predict} \begin{array}[b]{rrlcrl} 
C_{7+}^{}\,\mbox{\bf:} ~~~ &~ {-}4.4 \!\times\! 10^{-4} & < \hat\Delta_\mu < 11 \!\times\! 10^{-4} \,, & ~~~ & {-}4.1 \!\times\! 10^{-4} & < \hat\Delta_e^{} < 7.7 \!\times\! 10^{-4} \,,
\vspace{2pt} \\ &
{-}3.8 \!\times\! 10^{-4} & < \hat\Delta_\gamma < 7.2 \!\times\! 10^{-4} \,, & & {-}1.1 \!\times\! 10^{-4} & < \hat A_\gamma < {+}0.6 \!\times\! 10^{-4} \,,
\vspace{1ex} \\
C_{7-}^{}\,\mbox{\bf:} ~~~ & {-}1.7 \!\times\! 10^{-2} & < \hat\Delta_\mu < 3.7 \!\times\! 10^{-2} \,, & &
{-}2.9 \!\times\! 10^{-3} & < \hat\Delta_e^{} < 6.2 \!\times\! 10^{-3} \,,
\vspace{2pt} \\ &
{-}2.6 \!\times\! 10^{-3} & < \hat\Delta_\gamma < 5.5 \!\times\! 10^{-3} \,, & & {-}3.7 \!\times\! 10^{-3} & < \hat A_\gamma < 7.9 \!\times\! 10^{-3} \,,
\vspace{1ex} \\
C_{9+}^{}\,\mbox{\bf:} ~~~ & {-}1.2 \!\times\! 10^{-3} & < \hat\Delta_\mu < 1.2 \!\times\! 10^{-3} \,, & &
{-}3.8 \!\times\! 10^{-5} & < \hat\Delta_e^{} < {+}2.8 \!\times\! 10^{-5} \,, \vspace{1ex} \\
C_{9-}^{}\,\mbox{\bf:} ~~~ & {-}0.32 & < \hat\Delta_\mu < 0.32 \,, & & {-}0.13 & < \hat\Delta_e^{} < {+}0.12 \,. \end{array} ~
\end{align} 
For comparison, whereas the data on $\hat\Delta_\mu$ and $\hat\Delta_e$ are not yet available, the aforementioned search for $CP$ violation in \,$\Sigma^+ {\,\to\,} p\gamma$\, by BESIII produced~\cite{BESIII:2023fhs}
\begin{align}
\hat\Delta_\gamma^{\tt exp} & = 0.006 \pm 0.011_{\rm stat} \pm 0.004_{\rm syst} \,, &
\hat A_\gamma^{\tt exp} & = 0.095 \pm 0.087_{\rm stat} \pm 0.018_{\rm syst} \,, 
\end{align}
which are consistent with $CP$ invariance.
Evidently, these data are not yet precise enough to pose substantial tests for the $\hat\Delta_\gamma$ and $\hat A_\gamma$ predictions in eq.\,(\ref{predict}), but future efforts~\cite{Schonning:2023mge} by PANDA \cite{PANDA:2021ozp} and at the Super Tau Charm Facility \cite{Achasov:2023gey} can be anticipated to probe these quantities more stringently. 

It is instructive to look at how sizable Im$(\lambda_tC_{7\pm})$ and Im$(\lambda_tC_{9\pm})$ could get in specific NP models, which may be subject to additional constraints.
This we do in the next few subsections.

\subsection{Supersymmetric models\label{susy}}

In supersymmetric extensions of the SM with generic flavor couplings and minimal particle content, enlarged $C_{7\pm}$ could be brought about by one-loop diagrams involving gluinos and squarks, which remove the chirality suppression occurring in the SM.
The enhancement similarly modifies the accompanying gluonic-dipole interactions.
The relevant effective Hamiltonian can be written as \cite{Buras:1999da}
\begin{equation} \label{H_mo}
{\cal H}_{\rm eff}^{\gamma,g} \,=\, C_\gamma^+ Q_{\gamma}^+ + C_\gamma^- Q_{\gamma}^- + C_g^+ Q_g^+ + C_g^- Q_g^- \,+\, \rm H.c., ~~~ ~~
\end{equation}
which contains the complex Wilson coefficients $C_{\gamma,g}^\pm$  and the magnetic operators
\begin{align}
Q_\gamma^\pm & = \frac{-\tt e\,}{48\pi^2}\, \overline s  \big(P\!_R^{}\pm P\!_L^{}\big) \sigma_{\kappa\tau}^{} F^{\kappa\tau} d \,, & Q_g^\pm & = \frac{g_{\rm s}^{}}{32\pi^2}\, \overline s \big(P\!_R^{}\pm P\!_L^{}\big) \sigma_{\kappa\tau}^{} \lambda_a^{} G_a^{\kappa\tau} d \,, 
\end{align}
where $g_{\rm s}^{}$ symbolizes the strong coupling constant and $\lambda_a^{}G_a^{\kappa\tau}$ denotes a Gell-Mann 
\pagebreak 
matrix acting on color space times the gluon
field-strength tensor, with 
\,$a=1,2,...,8$\, being summed over. 
Obviously, $C_\gamma^\pm$ are just $C_{7\pm}^{}$ up to known factors.
In the so-called mass-insertion approximation \cite{Gabbiani:1996hi} the coefficients at the supersymmetry breaking scale are given by~\cite{Buras:1999da}
\begin{equation} \label{F0G0}
C_\gamma^\pm (m_{\tilde g}^{}) \,=\, \frac{\alpha_{\rm s}^{}(m_{\tilde g})\,\pi}{m_{\tilde g}^{}} \Big[ \big(\delta_{LR}^D\big)_{21} \pm \big(\delta_{LR}^D\big)_{12}^* \Big]
F_0^{} (x_{gq}) \,=\,
C_g^\pm (m_{\tilde g})\, \frac{F_0^{} (x_{gq})}{G_0^{} (x_{gq})} \,, ~~~
\end{equation}
where the $\delta$s are the parameters of the mass-insertion formalism, \,$x_{gq}^{}=m_{\tilde g}^2/m_{\tilde q}^2$,\, with $m_{\tilde g}$ and $m_{\tilde q}$ being the gluino and average squark masses,  and
$F_0$ and $G_0$ represent loop functions.

In illustrating the implications for our processes of interest, we pick \,$x_{gq}=1$\,  for definiteness, in which case
\,$F_0^{}(x_{gq}) = 2/9$\, and \,$G_0^{}(x_{gq}) = -5/18$.\,
For $m_{\tilde g}$ exceeding the top-quark mass, at the renormalization scale \,$\mu=1$ GeV\, we have \cite{Buras:1999da}
\begin{align}
C_\gamma^\pm(\mu) & \,=\, \tilde\eta^2 C_\gamma^\pm( m_{\tilde g}) + 8 \big( \tilde\eta^2 - \tilde\eta\big) C_g^\pm (m_{\tilde g}) \,=\, \bigg(\frac{36}{5} \tilde\eta - 8\bigg) \tilde\eta\, C_g^\pm(m_{\tilde g}) \,, ~~~ ~~~~
\nonumber \\
C_g^\pm(\mu) & \,=\, \tilde\eta\, C_g^\pm (m_{\tilde g}) \,,
\end{align}
where the last equality in eq.\,(\ref{F0G0}) has been incorporated into $C_\gamma^\pm (\mu)$ and
\begin{equation}
\tilde\eta \,=\, \bigg(\frac{\alpha_{\rm s}(m_{\tilde g})}{\alpha_{\rm s}(m_t)}\bigg)\raisebox{8pt}{$^{\!2/21}$}\, \bigg(\frac{\alpha_{\rm s}(m_t)}{\alpha_{\rm s}(m_b)}\bigg)\raisebox{8pt}{$^{\!2/23}$}\,
\bigg(\frac{\alpha_{\rm s}(m_b)}{\alpha_{\rm s}(m_c)}\bigg)\raisebox{8pt}{$^{\!2/25}$}\,
\bigg(\frac{\alpha_{\rm s}(m_c)}{\alpha_{\rm s}(\mu)}\bigg)\raisebox{8pt}{$^{\!2/27}$} . ~~~ ~~~~
\end{equation}
Choosing \,$m_{\tilde g}=m_{\tilde q}=3$ TeV,\, which is in line with the corresponding mass limits from collider  direct searches to date \cite{ParticleDataGroup:2024cfk}, then translates into \,$C_g^\pm (\mu) = 0.86\, C_g^\pm\big(m_{\tilde g})$\, and consequently \,$C_\gamma^\pm (\mu) = -1.8\, C_g^\pm(\mu)$.\,

There are restraints on Im\,$C_g^\pm(\mu)$ imposed by the measurements of $CP$ violation in neutral-kaon mixing and decay \,$K\to\pi\pi$,\, respectively.\footnote{More probes of Im\,$C_g^\pm$ could turn up from the quests for $CP$ violation in nonleptonic hyperon\linebreak decays \cite{Donoghue:1986hh,He:1995na,He:1999bv,Tandean:2003fr,He:2025wxn}, but the search results so far \cite{ParticleDataGroup:2024cfk,Zheng:2025tnz} are not yet sufficiently precise to test for it.}
From a very recent assessment in~ref.\,\cite{He:2025wxn}, we deduce the 90\%-CL bounds
\begin{align} \label{ImCg}
10^9\hspace{1pt} \big|{\rm Im}\hspace{1pt}C_g^+(\mu)\big| & < 7.7 \rm\;GeV^{-1} , &  
-8.3 {\rm\;GeV}^{-1} < 10^9\, {\rm Im}\hspace{1pt}C_g^-(\mu) & < +4.6 \rm\;GeV^{-1} \,. ~ 
\end{align}
Putting these together with the $C_\gamma$-$C_g$ relation specified above, we obtain
\begin{align} \label{ImCgamma}
10^8\hspace{1pt} \big|{\rm Im}\hspace{1pt}C_\gamma^+(\mu)\big| & < 1.4 \rm\;GeV^{-1} , &  
-0.8 {\rm\;GeV}^{-1} < 10^8\, {\rm Im}\hspace{1pt}C_\gamma^-(\mu) & < 1.5 \rm\;GeV^{-1} 
\end{align}
and hence
\begin{align} \label{ImC7+-}
\big|{\rm Im}\big(\lambda_t^{}C_{7+}^{}(\mu)\big)\big| & < 2.3\!\times\!10^{-3} , &  
-2.5\!\times\!10^{-3} < {\rm Im}\big(\lambda_t^{}C_{7-}^{}(\mu)\big) & < +1.4\!\times\!10^{-3} \,. ~  
\end{align} 
These ranges in eq.\,(\ref{ImC7+-}) are smaller than the corresponding model-independent ones 
\pagebreak 
quoted in eq.\,(\ref{indepImC7}) roughly by one and two orders of magnitude, respectively.
It is clear that this is caused by $C_\gamma^\pm$ being closely connected to $C_g^\pm$ in this NP scenario as well as by the severe restrictions on $C_g^\pm$ from kaon nonleptonic processes.
It can notwithstanding be said that the predicted asymmetries in eq.\,(\ref{predict}) due to $C_{7\pm}$ alone might yet be attainable in models where such a connection is absent.

\subsection{Scenarios with \boldmath$ds\nu\nu$ interactions\label{numodels}}

In some ultraviolet completions several of the operators with charged leptons are linked to the operators with SM neutrinos
\begin{align} \label{OLR}
{\cal O}_L^{ij} & \,=\, \frac{\lambda_t^{}{\tt e}^2G_{\rm F}^{}}
{2\sqrt2\,\pi^2}\, \overline s \gamma_\mu^{} P\!_L^{} d~ \overline{\nu_i^{}} \gamma^\mu P\!_L^{}\nu_j^{} \,, & {\cal O}_R^{ij} & \,=\, \frac{\lambda_t^{}{\tt e}^2G_{\rm F}^{}}
{2\sqrt2\,\pi^2}\, \overline s \gamma_\mu^{} P\!_R^{}\, d~ \overline{\nu_i^{}} \gamma^\mu P\!_L^{}\nu_j^{} \,. 
\end{align}
This could happen when the charged and neutral leptons involved belong to left-handed doublets in their interactions with other particles.
In such instances, the coefficients of the two sets of operators may be correlated, and this could entail additional, even stringent, constraints from decays with neutrinos.
The coefficients $C_{L,R}^{ij}$ of ${\cal O}_{L,R}^{ij}$ can be combined as~\,$C_{\nu\nu\pm}^{ij}\equiv C_L^{ij}\pm C_R^{ij}$.\,

In the presence of lepton-flavor-diagonal NP, the branching fractions of \,$K\to\pi\nu\bar\nu$\, including the SM contributions are 
\begin{align} \label{kpinunu} & \begin{array}[b]{rl} \displaystyle 
{\cal B}\big(K^+{\to\,}\pi^+\nu\bar\nu(\gamma)\big) \,= & \displaystyle \raisebox{4pt}{\footnotesize$\displaystyle
\sum_{\mbox{\scriptsize$j\!=\!e{,}\mu{,}\tau$}}$} \frac{\kappa_+}{3} (1+\Delta_{\rm EM}) \Bigg\{ \frac{1}{\lambda^{10}} \big[ {\rm Im}\big( \lambda_t^{} X_t^{} \!-\! \lambda_t^{} C^{jj}_{\nu\nu+} s_{\rm w}^2 \big) \big]^2 
\\ & \displaystyle \hspace{8ex} +~ \bigg[ \frac{P_c}{\lambda} {\rm Re}\,\lambda_c^{} + \frac{1}{\lambda^5} {\rm Re} \big( \lambda_t^{} X_t^{} \!-\! \lambda_t^{} C^{jj}_{\nu\nu+} s_{\rm w}^2 \big) \bigg]^2 \Bigg\} \,, \end{array}
\\ & 
{\cal B} \big(K_L{\,\to\,}\pi^0\nu\bar\nu\big) \,=\, \raisebox{4pt}{\footnotesize$\displaystyle
\sum_{\mbox{\scriptsize$j\!=\!e{,}\mu{,}\tau$}}$}\, \frac{\kappa_L}{3} \frac{\big[ {\rm Im}\big( \lambda_t^{} X_t^{} - \lambda_t^{} C^{jj}_{\nu\nu+} s_{\rm w}^2 \big) \big]^2}{\lambda^{10}} \,, 
\end{align}
where \,$\lambda=|V_{us}|$\, and the other constants are defined in the literature \cite{Buchalla:1995vs,Brod:2021hsj,ParticleDataGroup:2024cfk}.
In the case of \,$C^{jj}_{\nu\nu+}=0$,\, these turn into the SM predictions \cite{ParticleDataGroup:2024cfk}
\begin{align}
{\cal B}(K^+\to\pi^+\nu\bar\nu)_{\tt SM}^{} & \,=\, (8.2\pm 0.6)\times 10^{-11} \,, & 
\nonumber \\
{\cal B}(K_L \to\pi^0\nu\bar\nu)_{\tt SM}^{} & \,=\, (2.9\pm 0.3)\times 10^{-11} \,,
\end{align} 
where the uncertainties are parametric and reflect mostly those in the CKM angles.

Using the latest NA62 finding \,${\cal B}(K^+\to\pi^+\nu\overline\nu)_{\tt exp}  = \big(9.6^{+1.9}_{-1.8}\big) \times 10^{-11\,}$ \cite{Chang:2026vvx},  we can express  
\begin{align} \label{xBK2pinunu}
\frac{{\cal B}(K^+\to\pi^+\nu\bar\nu)_{\tt exp}-{\cal B}(K^+\to\pi^+\nu\bar\nu)_{\tt SM}}{{\cal B}(K^+\to\pi^+\nu\bar\nu)_{\tt SM}} & \,=\, 0.17_{-0.23}^{+0.24} \,,
\nonumber \\
{\cal B}(K^+\to\pi^+\nu\bar\nu)_{\tt exp}-{\cal B}(K^+\to\pi^+\nu\bar\nu)_{\tt SM}^{} & \,=\, \big(1.4^{+2.0}_{-1.9}\big)\times 10^{-11} \,. ~~~
\end{align}
Since the implied Grossman-Nir bound from the NA62 measurement is still below the 
\pagebreak 
KOTO limit \cite{ParticleDataGroup:2024cfk}, the best current restraint on Im$(\lambda_t C_{\nu\nu+})$ is supplied by \,$K^+\to\pi^+\nu\bar\nu$.\,  
Supposing that NP impacts just one neutrino flavor and is such that \,${\rm Re}(\lambda_t C_{\nu\nu+})=0$,\, in view of eqs.\,\,(\ref{kpinunu}) and (\ref{xBK2pinunu}) we can then extract the 90\%-CL range 
\begin{equation} \label{ImCnn}
-3.2\times 10^{-3} \,<\, {\rm Im}(\lambda_t C_{\nu\nu+}) \,<\, 5.0\times 10^{-3} \,. ~~~
\end{equation}
As for Im$(\lambda_t C_{\nu\nu-})$, it affects \,$K^+\to\pi^+\pi^0\nu\bar\nu$,\, but the constraint inferred from its data~\cite{ParticleDataGroup:2024cfk} is presently unimportant \cite{Tandean:2019tkm,Roy:2024avj}.\footnote{Both ${\cal O}_{L,R}^{ij}$ also influence rare hyperon decays with missing energy \cite{Tandean:2019tkm,Su:2019tjn}, which might be pursued intensively at future facilities \cite{Zhang:2026pyv}.}

\subsection{Scalar leptoquarks\label{LQ}}

These bosons from beyond the SM can alter rare semileptonic processes, as studied in refs.\,\cite{Davies:1990sc,Davidson:1993qk,Valencia:1994cj,Dorsner:2016wpm,Hiller:2016kry,Bobeth:2017ecx,Mandal:2019gff,Su:2019tjn} among others. In the nomenclature of ref.\,\cite{Dorsner:2016wpm}, the spinless leptoquarks (with their gauge ${\rm SU}(3)_{\rm color}\times{\rm SU}(2)_L\times{\rm U}(1)_Y$\, assignments) which possess couplings that conserve baryon and lepton numbers, respect SM gauge symmetries, and can contribute to \,$\Sigma^+ {\,\to\,} p\ell^+\ell^-$\, are \,$\widetilde{\mathscr S}_1\,\big(\bar3,1,4/3\big)$,\, $\mathscr R_2\,(3,2,7/6)$,\, $\widetilde{\mathscr R}_2\,(3,2,1/6)$,  and~\,$\mathscr S_3\,\big(\bar3,3,1/3\big)$.
In terms of their components, we have 
\begin{align}
\widetilde{\mathscr S}_1^{} = \tilde S{}_1^{4/3} , & & \mathscr R_2^{} = \Bigg( \begin{array}{c} R_2^{5/3} \vspace{3pt} \\ R_2^{2/3} \end{array} \Bigg) , & & \widetilde{\mathscr R}_2^{}  = \Bigg( \begin{array}{c} \tilde R{}_2^{2/3} \vspace{3pt} \\ \tilde R{}_2^{-1/3} \end{array} \Bigg) , & & \mathscr S_3^{} = \Bigg( \begin{array}{c} S_3^{1/3} ~~~~~
\sqrt2\, S_3^{4/3} \vspace{3pt} \\ \sqrt2\, S_3^{-2/3} ~~~ {-}S_3^{1/3} \end{array} \Bigg) ,
\end{align}
where each superscript refers to the electric charge.
The renormalizable Lagrangian pertaining to the fermionic interactions of these leptoquarks (LQs) has the form
\begin{equation} \label{Llq}
{\cal L}_{\tt LQ}^{} \,=\, {\tt Y}_{1,jy}^{\textsc{rr}}\, \overline{d_j^{\textsc c}}e_y^{}\widetilde{\mathscr S}_1^{} + {\tt Y}_{2,jy}^{\textsc{lr}}\, \overline{q_j^{}} e_y^{}\mathscr R_2^{} + {\tt Y}_{2,jy}^{\textsc{rl}}\, \overline{d_j^{}} \widetilde{\mathscr R}_2^{\textsc t} i\tau_2^{} l_y^{} + {\tt Y}_{3,jy}^{\textsc{ll}}\, \overline{q_j^{\textsc c}} i\tau_2^{} {\mathscr S}_3^{} l_y^{} \,+\, \rm H.c. ~
\end{equation}
Here the ${\tt Y}$s stand for 3$\times$3 Yukawa matrices whose elements are dimensionless free parameters which can be complex, $q_j^{}$ and $l_j$ $\big(d_j$ and $e_j\big)$ represent the left-handed doublets (right-handed singlets) of quarks and leptons, respectively, the superscript $\textsc c$ symbolizes charge conjugation, the family indices \,$j,y=1,2,3$\, are summed over, and $\tau_2$ is the second Pauli matrix.

Assumed to be heavy, the LQs in eq.\,(\ref{Llq}) can mediate at tree level the interactions of quarks with charged leptons described by the operators in eq.\,(\ref{Oj}), their nonzero coefficients being given by
\begin{align} \label{Cl}
\frac{\alpha_{\tt e}^{}\lambda_t^{}G_{\rm F}^{}}{\sqrt2\,\pi}\, C_9^\ell & \,=\,
\frac{-{\tt Y}_{2,1\ell}^{\textsc{lr}*}\, {\tt Y}_{2,2\ell}^{\textsc{lr}}}{4\, m_{\mathscr R_2}^2} + \frac{{\tt Y}_{3,2\ell}^{\textsc{ll}*}\, {\tt Y}_{3,1\ell}^{\textsc{ll}}}{2\, m_{\mathscr S_3}^2} \,,  
\nonumber \\
\frac{\alpha_{\tt e}^{}\lambda_t^{}G_{\rm F}^{}}{\sqrt2\,\pi}\, C_{9'}^\ell & \,=\, \frac{{\tt Y}_{1,2\ell}^{\textsc{rr}*}\, {\tt Y}_{1,1\ell}^{\textsc{rr}}}{4\, m_{\widetilde{\mathscr S}_1}^2} - \frac{{\tt Y}_{2,1\ell}^{\textsc{rl}*}\, {\tt Y}_{2,2\ell}^{\textsc{rl}}}{4\, m_{\widetilde{\mathscr R}_2}^2} \,,
\nonumber \\
\frac{\alpha_{\tt e}^{}\lambda_t^{}G_{\rm F}^{}}{\sqrt2\,\pi}\, C_{10}^\ell & \,=\, \frac{-{\tt Y}_{2,1\ell}^{\textsc{lr}*}\, {\tt Y}_{2,2\ell}^{\textsc{lr}}}{4\, m_{\mathscr R_2}^2} - \frac{{\tt Y}_{3,2\ell}^{\textsc{ll}*}\, {\tt Y}_{3,1\ell}^{\textsc{ll}}}{2\, m_{\mathscr S_3}^2} \,,  
\nonumber \\
\frac{\alpha_{\tt e}^{}\lambda_t^{}G_{\rm F}^{}}{\sqrt2\,\pi}\, C_{10'}^\ell & \,=\, \frac{{\tt Y}_{1,2\ell}^{\textsc{rr}*}\, {\tt Y}_{1,1\ell}^{\textsc{rr}}}{4\, m_{\widetilde{\mathscr S}_1}^2} + \frac{{\tt Y}_{2,1\ell}^{\textsc{rl}*}\, {\tt Y}_{2,2\ell}^{\textsc{rl}}}{4\, m_{\widetilde{\mathscr R}_2}^2} \,, & 
\end{align}
and $C_{7,7'}$ created at one-loop level and hence comparatively negligible.
The $\widetilde{\mathscr R}_2$ and $\mathscr S_3$ terms in eq.\,(\ref{Llq}) can also modify the quark-neutrino transitions characterized by~eq.\,(\ref{OLR}), generating the contributions
\begin{align} \label{CLCR}
\frac{\alpha_{\tt e}^{}\lambda_t^{}G_{\rm F}^{}}{\sqrt2\,\pi}\, C_L^{ij} & \,=\, \frac{{\tt Y}_{3,2i}^{\textsc{ll}*} {\tt\,Y}_{3,1j}^{\textsc{ll}}}{4\, m_{\mathscr S_3}^2} \,, &
\frac{\alpha_{\tt e}^{}\lambda_t^{}G_{\rm F}^{}}{\sqrt2\,\pi}\, C_R^{ij} & \,=\, \frac{-{\tt Y}_{2,1i}^{\textsc{rl}*} {\tt\,Y}_{2,2j}^{\textsc{rl}}}{4\, m_{\widetilde{\mathscr R}_2}^2} \,. ~~~ ~~
\end{align}
Clearly, \,$C_{9(9')}^\ell=C_{10(10')}^\ell$\, in the $\widetilde{\mathscr S}_1$ $(\mathscr R_2)$ model, but it has no direct link to $C_{L,R}^{ij}$, as the LQ is not coupled to a left-handed lepton doublet according to eq.\,(\ref{Llq}).
By contrast, \,$C_{9'}^\ell=-C_{10'}^\ell=C_R^{\ell\ell}$\, for $\widetilde{\mathscr R}_2$ and \,$C_9^\ell=-C_{10}^\ell=2C_L^{\ell\ell}$\, for $\mathscr S_3$.

Although ${\cal L}_{\tt LQ}$ can bring about changes to a variety of processes, the strongest restrictions on the imaginary parts of the products of Yukawa couplings $\big(Y_{s\ell}^*Y_{d\ell}^{}\big)$ listed in eqs.\,\,(\ref{Cl}) and (\ref{CLCR}), if only one of the LQs is present at a time, have turned out to come from \,$K_L\to\pi^0\ell^+\ell^-$\, and \,$K^+\to\pi^+\nu\bar\nu$\, measurements \cite{Bobeth:2017ecx,Mandal:2019gff}, as already indicated in sections \ref{modindep} and \ref{numodels}.
Thus, for each of the LQs, we can derive the allowed range of Im$\big(Y_{s\ell}^*Y_{d\ell}^{}\big)$ from the experimental limit on the \,$K_L\to\pi^0\ell^+\ell^-$\, rate, after implementing its formula as written down in refs.\,\cite{Roy:2024hqg,Mescia:2006jd,DAmbrosio:2022kvb} and incorporating $C_{9+}^\ell$ and $C_{10+}^\ell$ from eq.\,(\ref{Cl}) in terms of the Yukawa couplings.
We do likewise from eq.\,(\ref{ImCnn}) with $C_{\nu\nu+}^{\ell\ell}$ from eq.\,(\ref{CLCR}).
The results are collected in table \ref{ImY*Y-ranges}, where $m_{\tt LQ}$ denotes the LQ mass.
It is worth remarking that with, for example, \,$m_{\tt LQ}=2$ TeV,\, which is consistent with the LQ-mass limits from collider direct searches \cite{ParticleDataGroup:2024cfk}, the size of Im$\big(Y_{s\ell}^*Y_{d\ell}^{}\big)$ in this table is at most about \,$3\times10^{-3}$,\, which satisfies the perturbativity requirement for the Yukawa couplings.

\begin{table}[b]\bigskip\centering\footnotesize\setlength{\tabcolsep}{0pt}\begin{tabular}{|c|c|rl|rl|} \hline Process & \tt~LQ~ & \,Allowed range~ & in unit of $m\tt_{LQ}^2/TeV^2$\, & Resulting ranges of & \, Im$(\lambda_tC_{9\pm})\vphantom{|_|^|}$ \\ \hline \hline
$\begin{array}{l}K_L\to\\~\pi^0\mu^+\mu^-\end{array}$ & $\begin{array}{c} \widetilde{\mathscr S}_1 \vspace{2pt} \\ \mathscr R_2 \vspace{2pt} \\ \widetilde{\mathscr R}_2 \vspace{2pt} \\ \mathscr S_3 \end{array}_{\vphantom{|}}^{\vphantom{|}}$ & Im$\big(Y_{s\mu}^*Y_{d\mu}^{}\big)$\,=&$\left\{ \!\! \begin{array}{l} (-7.2, 7.0) \!\times\!10^{-4^{\vphantom{|}}} \vspace{2pt} \\ (-7.2, 7.0) \!\times\!10^{-4} \vspace{2pt} \\ \mbox{$(-8.0, 6.3)$} \!\times\!10^{-4} \vspace{2pt} \\ \mbox{$(-4.0, 3.2)$} \!\times\!10^{-4} \end{array} \right.$ & \,$\begin{array}{r} {\rm Im}\big(\lambda_t^{}C_{9+}^\mu\big) \!=\! {\rm- Im}\big(\lambda_t^{}C_{9-}^\mu\big) \vspace{2pt} \\ {\rm Im}\big(\lambda_t^{}C_{9+}^\mu\big) \!=\! {\rm Im}\big(\lambda_t^{}C_{9-}^\mu\big) \vspace{2pt} \\ {\rm Im}\big(\lambda_t^{}C_{9+}^\mu\big) \!=\! {\rm-Im}\big(\lambda_t^{}C_{9-}^\mu\big) \vspace{2pt} \\ {\rm Im}\big(\lambda_t^{}C_{9+}^\mu\big) \!=\! {\rm Im}\big(\lambda_t^{}C_{9-}^\mu\big) \end{array}$ & $\!\!\begin{array}{l} =\! (-8.8,8.5) \!\times\!10^{-3} \vspace{2pt} \\ =\! (-8.8,8.5) \!\times\!10^{-3} \vspace{2pt} \\ =\! (-9.7,7.7) \!\times\!10^{-3} \vspace{2pt} \\ =\! (-9.7,7.7) \!\times\!10^{-3} \end{array}$\, \\ \hline
$\begin{array}{l}K_L\to\\~\pi^0e^+e^-\end{array}$ & $\begin{array}{c} \widetilde{\mathscr S}_1 \vspace{1pt} \\ \mathscr R_2 \vspace{1pt} \\ \widetilde{\mathscr R}_2 \vspace{1pt} \\ \mathscr S_3 \end{array}_{\vphantom{|}}^{\vphantom{|}}$ & Im$\big(Y_{se}^*Y_{de}^{}\big)$\,=&$\left\{ \!\! \begin{array}{l} (-4.2, 3.1) \!\times\!10^{-4^{\vphantom{|}}}
\vspace{1pt} \\ (-4.2, 3.1) \!\times\!10^{-4}
\vspace{1pt} \\ \mbox{$(-4.8, 2.7)$} \!\times\!10^{-4} \vspace{1pt} \\
\mbox{$(-2.4, 1.4)$} \!\times\!10^{-4} \end{array} \right.$ & $\begin{array}{r} {\rm Im}\big(\lambda_t^{}C_{9+}^e\big) \!=\! {\rm- Im}\big(\lambda_t^{}C_{9-}^e\big) \vspace{1.9pt} \\ {\rm Im}\big(\lambda_t^{}C_{9+}^e\big) \!=\! {\rm Im}\big(\lambda_t^{}C_{9-}^e\big) \vspace{1.9pt} \\ {\rm Im}\big(\lambda_t^{}C_{9+}^e\big) \!=\! {\rm-Im}\big(\lambda_t^{}C_{9-}^e\big) \vspace{1.9pt} \\ {\rm Im}\big(\lambda_t^{}C_{9+}^e\big) \!=\! {\rm Im}\big(\lambda_t^{}C_{9-}^e\big) \end{array}$ & $\!\!\begin{array}{l} =\! (-5.1,3.8) \!\times\!10^{-3} \vspace{1pt} \\ =\! (-5.1,3.8) \!\times\!10^{-3} \vspace{1pt} \\ =\! (-5.8,3.3) \!\times\!10^{-3} \vspace{1pt} \\ =\! (-5.8,3.3) \!\times\!10^{-3} \end{array}$ \\ \hline
$\begin{array}{l}K^+\to\\~~\pi^+\nu\bar\nu\end{array}$ & $\begin{array}{c} \widetilde{\mathscr R}_2 \vspace{1pt} \\ \mathscr S_3 \end{array}_{\vphantom{|}}^{\vphantom{|}}$ & Im$\big(Y_{s\ell}^*Y_{d\ell}^{}\big) \vphantom{\Big|}$\,=&$\left\{ \!\! \begin{array}{l} (-2.7, 4.2) \!\times\!10^{-4^{\vphantom{|}}} \\ (-2.7, 4.2) \!\times\!10^{-4^{\vphantom{|}}} \end{array} \right.$ & $\begin{array}{r} {\rm Im}\big(\lambda_t^{}C_{9+}^\ell\big) \!=\! {\rm-Im}\big(\lambda_t^{}C_{9-}^\ell\big) \vspace{1.9pt} \\ {\rm Im}\big(\lambda_t^{}C_{9+}^\ell\big) \!=\! {\rm Im}\big(\lambda_t^{}C_{9-}^\ell\big) \end{array}$ & $\!\!\begin{array}{l} =\! (-3.3, 5.1) \!\times\!10^{-3} \vspace{1pt} \\ =\! (-6.6, 10) \!\times\!10^{-3} \end{array}$ \\ \hline
\end{tabular}\caption{The ranges of the imaginary parts of \,$Y_{s\ell}^*Y_{d\ell}^{}$\, permitted by $K_L{\to}\hspace{1pt}\pi^0\ell^+\ell^-$  and  $K^+{\to}\hspace{1pt}\pi^+\nu\bar\nu$  data and belonging to the $\widetilde{\mathscr S}_1$, $\mathscr R_2$, $\widetilde{\mathscr R}_2$, and $\mathscr S_3$ leptoquarks as dictated by eq.\,(\ref{Llq}) if only one of them exists at a time.
The last column displays the Im$\big(\lambda_tC_{9\pm}^\ell\big)$ intervals calculated from the corresponding Im$\big(Y_{s\ell}^*Y_{d\ell}^{}\big)$ entries with the aid of eq.\,(\ref{Cl}).\label{ImY*Y-ranges}} 
\end{table}

This table reveals that for $\widetilde{\mathscr R}_2$ and $\mathscr S_3$ the restraints on Im$\big(Y_{s\ell}^*Y_{d\ell}^{}\big)$ from \,$K^+\to\pi^+\nu\bar\nu$\, tend to be slightly stricter than the ones from \,$K_L\to\pi^0\ell^+\ell^-$,\, more so in the \,$\ell=\mu$\, case.
From the last column we learn that none of the individual LQ interactions in eq.\,(\ref{Llq}) can fully saturate the model-independent bounds on Im$\big(\lambda_t^{}C_{9+}^\ell\big)$ in eq.\,(\ref{ImC9+-ranges}), as the Yukawa couplings also enter $C_{10+}^\ell$ which occurs in the \,$K_L\to\pi^0\ell^+\ell^-$\, amplitude. 
Dropping $C_{10+}^\ell$ would instead imply, for instance, \,Im$\big(Y_{s\mu}^*Y_{d\mu}^{}\big) = (-1.5, 1.1) \!\times\! 10^{-3}$\, and \,Im$\big(Y_{se}^*Y_{de}^{}\big) = (-6.9, 3.8) \!\times\! 10^{-4}$\, for both $\widetilde{\mathscr S}_1$ and $\mathscr R_2$. 
However, the Im$\big(\lambda_t^{}C_{9+}^\ell\big)$ numbers in table\,\,\ref{ImY*Y-ranges} are evidently within factors of two from those in eq.\,(\ref{ImC9+-ranges}). 
By contrast, the Im$\big(\lambda_t^{}C_{9-}^\ell\big)$ limits in eq.\,(\ref{ImC9--ranges}) are considerably beyond the reach of any one of these simple LQ models.
Nevertheless, this does not rule out that there might be other, more complicated scenarios of NP in which eq.\,(\ref{ImC9--ranges}) could be saturated.

Lastly, we note that, if nonvanishing $Y_{s\mu,d\mu}$ and $Y_{se,de}$ associated with each of the LQs appear together, there are severe, extra constraints from \,$K_L\to e^\pm\mu^\mp$\, data~\cite{Mandal:2019gff}.
This is to be kept in mind when reading the corresponding \,$\ell=\mu$\, and \,$\ell=e$\, entries in table\,\,\ref{ImY*Y-ranges}.
Moreover, since eq.\,(\ref{ImCnn}) presupposes that just one of the lepton flavors is impacted by~NP, it is to be understood that for $\widetilde{\mathscr R}_2$ and $\mathscr S_3$ the \,$\ell=\mu$\, and \,$\ell=e$\, intervals in this table do not apply simultaneously.

\section{Conclusions\label{concl}}

Motivated by the recent LHCb observation of \,$\Sigma^+ {\,\to\,} p\mu^+\mu^-$,\, which is in good agreement with the prediction of the SM, we have explored the possibility that this transition might exhibit $CP$ violation that is sufficiently substantial to be detectable in near-future experiments.
We demonstrate in particular that the dominant long-distance portion of the decay amplitude within the SM provides large absorptive phases that could drive significant\linebreak$CP$-violation through interference with potential new-physics contributions.
Complementarily, we perform analogous investigations concerning the dielectron mode, \,$\Sigma^+ {\,\to\,} pe^+e^-$,\, and the related weak radiative one, \,$\Sigma^+ {\,\to\,} p\gamma$.\,

In numerical work we determine model-independent bounds on the Wilson coefficients of the pertinent quark operators that may originate from beyond the SM and also look at how additional requisites could arise within specific models of new physics.
Our analysis shows that some of the coefficients are not yet well-restricted by kaon and hyperon data to date, if a model-independent approach is taken.
In table\,\,\ref{tab:placeholder} we summarize our findings of the $CP$-violating asymmetries resulting from the coefficients that have been evaluated one at a time.
We see that the effects of $C_{7-}$ and $C_{9-}$, which are poorly constrained at the moment, especially the latter, could give rise to $\hat\Delta_\mu$ magnitudes of up to tens of percent, far exceeding the SM expectation.
This can be regarded as a wide window of new physics and would therefore offer LHCb, which may be able achieve precision for $\hat\Delta_\mu$ of order several percent in future runs, opportunities to discover in it signs from beyond the SM or, if not, improve upon the existing limits on Im$(\lambda_tC_{7-})$ and Im$(\lambda_tC_{9-})$.
Upcoming quests for $CP$ violation in \,$\Sigma^+ {\,\to\,} pe^+e^-$,\, as well as in \,$\Sigma^+ {\,\to\,} p\gamma$,\, by LHCb and other collaborations may be anticipated to produce similarly interesting outcomes.

\vspace{3ex} 

\begin{table}[t]\centering\setlength{\tabcolsep}{0pt}\begin{tabular}{|c||c|c|c|c|} \hline\footnotesize
~Source~ & $\hat\Delta_\mu\vphantom{|_|^{|^|}}$ & $\hat\Delta_e$ & $\hat\Delta_\gamma$ & $\hat A_\gamma$ \\ \hline
SM & ~$(1.1\pm0.2) \!\times\! 10^{-4}\vphantom{\sum_|^{|^|}}$~ & ~$(4.5\pm0.4) \!\times\! 10^{-6}$~ & ~$(1.1\pm0.3) \!\times\! 10^{-7}$~ & ~$(-2.7\pm0.9) \!\times\! 10^{-7}$~ \\
$C_{7+}$ & $(-4.4, 11) \!\times\! 10^{-4}\vphantom{\sum_|^|}$ & $(-4.1, 7.7) \!\times\! 10^{-4}$ & $(-3.8, 7.2) \!\times\! 10^{-4}$ & $(-1.1, 0.6) \!\times\! 10^{-4}$ \\
$C_{7-}$ & $(-1.7, 3.7) \!\times\! 10^{-2}\vphantom{\sum_|^|}$ & $(-2.9, 6.2) \!\times\! 10^{-3}$ & $(-2.6, 5.5) \!\times\! 10^{-3}$ & $(-3.7, 7.9) \!\times\! 10^{-3}$ \\
$C_{9+}$ & $(-1.2, 1.2) \!\times\! 10^{-3}\vphantom{\sum_|^|}$ & $(-3.8, 2.8) \!\times\! 10^{-5}$ & $-$ & $-$ \\
$C_{9-}$ & $(-0.32, 0.32)\vphantom{\sum_|^|}$ & $(-0.13, 0.12)$ & $-$ & $-$ \\ \hline
\end{tabular}
\caption{The values of the $CP$-violating observables of \,$\Sigma^+ {\,\to\,} p\ell^+\ell^-$ $\big(\hat\Delta_\ell\big)$, \,$\ell=\mu,e$,\, and $\Sigma^+ {\,\to\,} p\gamma$ $\big(\hat\Delta_\gamma$ and $\hat A_\gamma\big)$ that can be induced by the SM, from eqs.\,\,(\ref{sm-Deltal}) and (\ref{sm-Deltag}), and by the SM in conjunction with possible NP encoded in the different Wilson coefficients, taken one at a~time and extracted model-independently, within their respective current 90\%-CL intervals, from eq.\,(\ref{predict}).\label{tab:placeholder}} 
\end{table} 

\appendix
\section{Branching fractions in the SM\label{smrates}}

As elaborated in refs.\,\,\cite{He:2005yn,He:2018yzu,Roy:2024hqg}, the computation of the rate of \,$\Sigma^+ {\,\to\,} p\ell^+\ell^-$\, needs input from the data \cite{Gershwin:1969fpe,Manz:1980td,Kobayashi:1987yv,E761:1992atm,BESIII:2023fhs,BESIII:2025rgd} on the decay asymmetry parameter $\alpha$ and branching fraction of \,$\Sigma^+ {\,\to\,} p\gamma$\, and \,$\Sigma^+ {\,\to\,} p\pi^0,n\pi^+$.\,
Following ref.\,\cite{Roy:2024hqg}, we employ 
\begin{align} \label{ag,Bg}
\alpha_{\Sigma^+\to p\gamma}^{\tt exp} & \,=\, -0.694 \pm 0.047 \,, & {\cal B}(\Sigma^+ {\to\,} p\gamma)_{\tt exp} & \,=\, (9.96 \pm 0.28) \times 10^{-4} \,.
\end{align}
In the nonleptonic case, we have 
\begin{align} \label{a-exp}
\alpha_{\Sigma^+\to p\pi^0}^{\tt exp} & \,=\, -0.9868 \pm 0.0019 \,, & \alpha_{\Sigma^+\to n\pi^+}^{\tt exp} & \,=\, 0.0514 \pm 0.0029 \,, ~
\end{align}
after averaging the known empirical results \cite{ParticleDataGroup:2024cfk,
BESIII:2020fqg,BESIII:2023sgt,BESIII:2025jxt}, including the antihyperon ones if available.
Furthermore, we take into account the new BESIII measurements \cite{BESIII:2025rgd} 
\begin{align} \label{B-exp}
{\cal B}(\Sigma^+ {\,\to\,} p\pi^0)_{\tt exp} & = (49.79 \pm 0.23)\% \,, &  
{\cal B}(\Sigma^+ {\,\to\,} n\pi^+)_{\tt exp} & = (49.87 \pm 0.30)\% \,, \,
\end{align}
which differ by 4.4$\sigma$ and 3.4$\sigma$ from the corresponding PDG values \cite{ParticleDataGroup:2024cfk}.
From eqs.\,\,(\ref{a-exp}) and (\ref{B-exp}), we obtain  
\begin{align} \label{ABdata}
A_{p\pi^0} & \,=\, -1.385 \pm 0.008 \,, & B_{p\pi^0} & \,=\, 11.80 \pm 0.09 \,, 
\nonumber \\ 
A_{n\pi^+} & \,=\,  0.047 \pm 0.003 \,, & B_{n\pi^+} & \,=\, 18.84 \pm 0.07 &
\end{align}
for the parameters defined in the amplitude \,${\cal M}_{\Sigma^+\to N\pi} = iG_{\rm F}^{}m_{\pi^+}^2\, \bar u_N (A_{N\pi} - \gamma_5B_{N\pi}) u_\Sigma$,\linebreak the small phases of $A_{N\pi}$ and $B_{N\pi}$ having been ignored.
With eq.\,(\ref{ABdata}), as well as\linebreak $f\!_\pi^{}=(92.07\pm0.85)$\,MeV\, for the pion decay constant \cite{ParticleDataGroup:2024cfk}, we carry out the procedure detailed in ref.\,\cite{He:2005yn} to  update the imaginary parts of the form factors $a$,\,$b$,\,$c$, and $d$. 
We can then get
\begin{align} 
{\rm Im}\,a(0) & \,=\, (2.954 \pm 0.045)\rm\,MeV \,, & {\rm Im}\,b(0) & \,=\, (-1.745 \pm 0.030)\rm\,MeV \,, 
\end{align}
which in light of eq.\,(\ref{ag,Bg}) translate into
\begin{align} 
{\rm Re}\,a & \,=\, (12.13 \pm 0.25)\rm\,MeV \,, & {\rm Re}\,b & \,=\, (-4.77 \pm 0.42)\rm\,MeV ~~~ ~
\end{align}
for the relativistic solution 4, the short-distance components having been neglected. 
As before \cite{He:2005yn}, we estimate Re\,$c$ and Re\,$d$ assuming vector-meson dominance.
With all this information, we arrive at the predictions
\begin{align} \label{B(Sp2pll)sm}
\mathscr B_{\mu\mu}^{\tt SM}  & \,=\, {\cal B}(\Sigma^+ {\,\to\,} p\mu^+\mu^-)_{\tt SM} \,=\, (1.2 \pm 0.1) \times 10^{-8} \,, 
\nonumber \\ 
\mathscr B_{ee}^{\tt SM} & \,=\, {\cal B}(\Sigma^+ {\,\to\,} pe^+e^-)_{\tt SM} \,=\, (7.4 \pm 0.3) \times 10^{-6} \,, &
\end{align}
which are virtually the same as their counterparts reported in ref.\,\cite{Roy:2024hqg}.

\acknowledgments

We thank Francesco Dettori, Cong Geng, and Xiaorong Zhou for useful communications.
J.T. thanks the Tsung-Dao Lee Institute, Shanghai Jiao Tong University, for kind hospitality and support during different stages of this research.
X.-G.H. was supported by the Fundamental Research Funds for the Central Universities, by the National Natural Science Foundation of the People's Republic of China (Nos. 12090064, 12375088, and W2441004).
The work of G.V. was supported in part by an Australian Research Council Discovery Project.

\bibliographystyle{jhep.bst}
\bibliography{jhep_refs.bib}

\end{document}